\documentclass[aps, superscriptaddress, longbibliography, notitlepage, nofootinbib, twocolumn]{revtex4-1}
\usepackage{amsmath,amssymb}
\usepackage{graphicx}
\usepackage{hyperref}
\usepackage{enumitem}
\usepackage{color}
\usepackage{tikzit}
\usepackage[bottom]{footmisc}
\begin{document}

\title{Dynamics of hot random hyperbolic graphs}

\author{Fragkiskos Papadopoulos}\email{f.papadopoulos@cut.ac.cy}
\affiliation{Department of Electrical Engineering, Computer Engineering and Informatics, Cyprus University of Technology, 3036 Limassol, Cyprus}
\author{Sofoclis Zambirinis}
\affiliation{Department of Electrical Engineering, Computer Engineering and Informatics, Cyprus University of Technology, 3036 Limassol, Cyprus}

\date{\today}

\begin{abstract}
We derive the most basic dynamical properties of random hyperbolic graphs (the distributions of contact and intercontact durations) in the hot regime (network temperature $T > 1$). We show that for sufficiently large networks the contact distribution decays as a power law with exponent $2+T > 3$ for durations $t > T$, while for $t < T$ it exhibits exponential-like decays. This result holds irrespective of the expected degree distribution, as long as it has a finite $T^{\text{th}}$ moment. Otherwise, the contact distribution depends on the expected degree distribution and we show that if the latter is a power law with exponent $\gamma \in (2, T+1]$, then the former decays as a power law with exponent $\gamma+1 > 3$. On the other hand, the intercontact distribution exhibits power-law decays with exponent $2-T \in (0, 1)$ for $T \in (1,2)$, while for $T > 2$ it displays linear decays with a slope that depends on the observation interval. This result holds irrespective of the expected degree distribution as long as it has a finite $T^{\text{th}}$ moment if $T \in (1,2)$, or a finite second moment if $T > 2$. Otherwise, the intercontact distribution depends on the expected degree distribution and if the latter is a power law with exponent $\gamma \in (2, 3)$, then the former decays as a power law with exponent $3-\gamma \in (0,1)$. Thus, hot random hyperbolic graphs can give rise to contact and intercontact distributions that both decay as power laws. These power laws however are unrealistic for the case of the intercontact distribution, as their exponent is always less than one. These results mean that hot random hyperbolic graphs are not adequate for modeling real temporal networks, in stark contrast to cold random hyperbolic graphs ($T < 1$).  Since the configuration model emerges at $T \to \infty$, these results also suggest that this is not an adequate null temporal network model.
\end{abstract}

\maketitle

\section{Introduction}

It has been shown that random hyperbolic graphs (RHGs) are adequate models for real-world complex networks, as they naturally and simultaneously possess many of their common structural characteristics, including heterogeneous degree distributions, strong clustering and the small-world property, cf.~\cite{Krioukov2009, Krioukov2010, Panagiotou2012, cliques2015,  boguna2020, fountoulakis2021}. The vast majority of existing work has focused on structural properties of \emph{network snapshots}, i.e., of \emph{static graphs}, where it has been shown that RHGs are adequate models only in the so-called ``cold regime", where the \emph{network temperature} $T$ in the model takes values between zero and one. This is because only when $T \in [0, 1)$ RHGs can have strong clustering, as observed in real systems. In particular, clustering in RHGs is maximized at $T = 0$, it decreases to zero with $T \in [0, 1)$, and it is asymptotically zero for $T >1$~\cite{Krioukov2010}.\footnote{By \emph{asymptotically} or \emph{in the thermodynamic limit}, we mean \emph{as the number of nodes $N$ tends to infinity}.}  At $T=1$ there is a phase transition in RHGs, and for $T > 1$ RHGs are in the ``hot regime"~\cite{Krioukov2010}.

Motivated by the common dynamical properties observed across different human proximity networks, such as the broad distributions of contact and intercontact durations~\cite{BarratOverview2015, StarniniDevices2017, chaintreau_paper, thomas_paper}, the basic \emph{dynamical properties} of cold RHGs have been recently analyzed in Ref.~\cite{Papadopoulos2019}. In particular, it has been shown that sequences of independent network realizations constructed by the RHG model in the cold regime exhibit power-law contact and intercontact distributions, with corresponding exponents $2+T \in (2, 3)$ and $2-T \in(1, 2)$, irrespective of the distribution of expected node degrees~\cite{Papadopoulos2019}. These distributions are compatible with contact and intercontact distributions found in real human proximity networks, which roughly follow power laws (with or without exponential cutoffs)~\cite{hui_paper, chaintreau_paper, thomas_paper, StarniniDevices2017, HighSchoolData2}. Further, the exponents of these distributions in cold RHGs lie within the ranges observed in real systems. In particular, studies have reported power-law exponents larger than or close to two for the contact distribution~\cite{Scherrer2008, SPcontactexp}, and between one and two for the intercontact  distribution~\cite{hui_paper, chaintreau_paper, Partners2011, HighSchoolData2}. In general, cold RHGs have been found adequate for realistically modeling human proximity networks, as also corroborated by the observation that epidemic and rumor spreading processes perform remarkably similar in real and modeled systems~\cite{Papadopoulos2019}. 

Snapshots of human proximity networks are often very sparse, consisting of a small number of interacting (i.e., non-zero degree) nodes, cf.~\cite{Papadopoulos2019}. Because of their extreme sparsity, clustering in such snapshots is often close to zero. Cold RHGs can create such sparse snapshots with almost zero clustering due to their sparsity~\cite{Papadopoulos2019}. However, RHGs in the hot regime can also create such zero-clustering snapshots. This raises the question of whether there can be instances where hot RHGs can also be adequate models for such networks.  Motivated by this question, here we derive the contact and intercontact distributions in temporal hot RHGs. We show that asymptotically the contact distribution decays as a power law with exponent $2+T > 3$ for durations $t > T$, while for $t < T$ it exhibits exponential-like decays. This result holds irrespective of the expected degree distribution, as long as it has a finite $T^{\textnormal{th}}$ moment. Otherwise, the contact distribution depends on the expected degree distribution and we show that if the latter is a power law with exponent $\gamma \in (2, T+1]$, then the former decays as a power law with exponent $\gamma+1 > 3$. On the other hand, the  intercontact distribution exhibits power-law decays with exponent $2-T \in (0, 1)$ for $T \in (1,2)$, while for $T > 2$ it displays linear decays with a slope that depends on the observation interval. This result holds irrespective of the expected degree distribution as long as it has a finite $T^{\textnormal{th}}$ moment if $T \in (1,2)$, or a finite $2^{\textnormal{nd}}$ moment if $T > 2$. Otherwise, the intercontact distribution depends on the expected degree distribution and we show that if the latter is a power law with exponent $\gamma \in (2, 3)$, then the former decays as a power law with exponent $3-\gamma \in (0,1)$. 

Thus, hot RHGs can give rise to contact and intercontact distributions that both decay as power laws. These power laws have always exponents greater than three in the case of the contact distribution, and less than one in the case of the intercontact distribution. Power-law intercontact distributions with exponents less than one are non-normalizable (in a sense that will become precise in Sec.~\ref{sec:intercontact_dist}), and have not been observed in real systems. Therefore, our results suggest that there are no instances where hot RHGs can be adequate models for real temporal networks. This result applies also to the configuration model~\cite{ChungLu2002, Newman2004} that emerges at $T \to \infty$~\cite{Krioukov2010}. We note that even though the configuration model has been successfully used as a \emph{null} model in static networks, cf.~\cite{Voitalov2020}, our results suggest that it cannot be used as a null model in temporal networks if one wants to have the intercontact distribution realistically fixed. Our findings are in agreement with earlier results suggesting that only cold RHGs are adequate models for real networks, as they can reproduce both the amount of clustering in network snapshots, as well as the abundance of recurrent components in temporal networks~\cite{Papadopoulos2019, Flores2018}. 
Here, we reach the same conclusion regarding the realism of hot RHGs but from a completely different path---that of (inter)contact distributions.

We note that the node coordinates in RHGs abstract the popularities and similarities of the nodes~\cite{Papadopoulos2012}.  As mentioned, realistic power-law (inter)contact distributions emerge in cold RHGs irrespective of the distribution of the expected node degrees (popularities)~\cite{Papadopoulos2019}. This finding suggests that the origin of broad (inter)contact distributions in real systems is primarily due to node similarities, instead of popularities. Our results support this hypothesis. In particular, even though we prove that power-law intercontact distributions can also emerge purely due to the heterogeneity of expected node degrees, we find that these distributions in this case have unrealistic exponents. Thus, we provide analytical support to the idea that the emergence of broad (inter)contact distributions in real systems is due to node similarities. As the (inter)contact distributions critically affect the dynamics of spreading processes~\cite{vazquez2007, timo2009, machens2013, gauvin2013}, this observation could aid epidemic control strategies, as it implies that for altering the (inter)contact distributions in real systems one needs to target the (inter)contact durations between similar nodes instead of the (inter)contact durations of popular nodes.

The rest of the paper is organized as follows. In the next section we provide an overview of RHGs starting from their equivalent formulation, i.e., the $\mathbb{S}^1$ model~\cite{Krioukov2010, Serrano2008}, focusing on the hot regime. We also overview the dynamic-$\mathbb{S}^1$ model from~\cite{Papadopoulos2019} and give the precise definitions of the contact and intercontact distributions. In Secs.~\ref{sec:contact_dist} and~\ref{sec:intercontact_dist}, we analyze the contact and intercontact distributions in hot RHGs and in two paradigmatic infinite-temperature limits (the configuration model with power-law distributed expected degrees and classical random graphs~\cite{SoRa51}). We conclude the paper with discussion in Sec.~\ref{sec:conclusion}.

\section{Preliminaries}
\label{sec:preliminaries}

\subsection{$\mathbb{S}^{1}$ model/RHGs in the hot regime}
\label{sec:S1}

In the $\mathbb{S}^{1}$ model~\cite{Krioukov2010} each node has hidden variables $\kappa, \theta$. The hidden variable $\kappa$ determines the node's expected degree in the resulting network. The hidden variable $\theta$ is the angular similarity coordinate of the node on a circle of radius $R=N/2\pi$, where $N$ is the total number of nodes. To construct a network with the $\mathbb{S}^{1}$-model that has size $N$, average node degree $\bar{k}$, and temperature $T > 0$, we perform the following steps according to the \textbf{$\mathbb{S}^{1}$-model definition}:
\begin{enumerate}
\item[(1)] coordinate assignment: for each node $i=1, 2,\ldots,N$, sample its angular coordinate $\theta_i$ uniformly at random from $[0, 2\pi]$, and its degree variable $\kappa_i$ from a probability density function (PDF) $\rho(\kappa)$;
\item[(2)] creation of edges: connect every pair of nodes $i, j$ with the Fermi-Dirac connection probability
\begin{align}
\label{eq:p_s1}
p(\chi_{ij})=\frac{1}{1+\chi_{ij}^{1/T}}.
\end{align}
\end{enumerate}
In the last expression, $\chi_{ij}$ is the effective distance between nodes $i$ and $j$
\begin{equation}
\label{eq:chi}
\chi_{ij} = \frac{R \Delta\theta_{ij}}{\mu \kappa_i \kappa_j},
\end{equation}
where $\Delta \theta_{ij}=\pi - | \pi -|\theta_i - \theta_j||$ is the similarity distance between nodes $i$ and $j$. We note that since $\theta$ is uniformly distributed on $[0, 2\pi]$, the PDF of $\Delta \theta$ is the uniform PDF on $[0, \pi]$, $f(\Delta \theta)=1/\pi$. 

Parameter $\mu$ in~(\ref{eq:chi}) is derived from the condition that the expected degree in the network is indeed $\bar{k}$. In the hot regime ($T > 1$), and for sparse networks ($N \gg \bar{k}$), we have
\begin{equation}
\label{eq:mu}
\mu=\frac{\bar{k}^T (1-1/T)^T}{2 N^{T-1} (\overline{\kappa^{1/T}})^{2T}},
\end{equation}
where $\overline{\kappa^{1/T}} = \int \kappa^{1/T} \rho(\kappa) \mathrm{d} \kappa$. Further, the expected degree of a node with hidden variable $\kappa$ is
\begin{equation}
\label{eq:kappa}
\bar{k}(\kappa)= \frac{\bar{k}}{\overline{\kappa^{1/T}}}\kappa^{1/T}.
\end{equation}

The $\mathbb{S}^{1}$ model is equivalent to RHGs, i.e., to the $\mathbb{H}^{2}$ model~\cite{Krioukov2010}, after transforming the degree variables $\kappa$ to radial coordinates $r$ via
\begin{equation}
\label{eq:r_map}
r=\hat{R}-2 \ln{\frac{\kappa}{\kappa_0}},
\end{equation}
where $\kappa_0$ is the smallest $\kappa$, while $\hat{R}$ is the radius of the hyperbolic disk where all nodes reside. In hot RHGs, $\hat{R}=2T \ln{(N/\nu)}$, where $\nu=(\pi \mu \kappa_0^2)^{1/T} N^{1-1/T}$~\cite{Krioukov2010}. After this change of variables, the effective distance in~(\ref{eq:chi}) becomes $\chi_{ij}=e^{(x_{ij}-\hat{R})/2}$, where
\begin{equation}
\label{eq:xij}
 x_{ij}=r_i+r_j+2 \ln{\frac{\Delta\theta_{ij}}{2}}.
 \end{equation}
Eq.~(\ref{eq:xij}) gives approximately the hyperbolic distance between nodes $i$ and $j$. See Ref.~\cite{Krioukov2010} for further details.  

Instead of working directly with the  $\mathbb{H}^{2}$ model, we will be working with the more convenient $\mathbb{S}^{1}$ model and in particular with its alternative formulation that we describe next.  

\subsection{Alternative formulation of the $\mathbb{S}^{1}$ model that uses expected degrees}

Instead of sampling and fixing for each node $i$ its hidden degree variable $\kappa_i$ in step~(1) of the $\mathbb{S}^{1}$ model definition, we can instead sample and fix its expected degree $k_i \equiv \bar{k}(\kappa_i)$. Requiring that~(\ref{eq:kappa}) holds, the corresponding $\kappa_i$ is
\begin{equation}
\kappa_i=\left(\frac{\overline{\kappa^{1/T}}}{\bar{k}}\right)^T k_i^T.
\end{equation}
Substituting the $\kappa_i$ and $\kappa_j$ in~(\ref{eq:chi}) with their above expression, we can re-write the effective distance between two nodes $i$ and $j$ as
\begin{equation}
\label{eq:chi_alt}
\chi_{ij} = \frac{R \Delta\theta_{ij}}{\tilde{\mu} k_i^T k_j^T},
\end{equation}
where 
\begin{equation}
\tilde{\mu} = \frac{(1-1/T)^T}{2 N^{T-1} \bar{k}^T}.
\end{equation}
Further, the connection probability in~(\ref{eq:p_s1}) can be re-written as
\begin{equation}
\label{eq:p_s1_alt}
p(k_i, k_j, \Delta \theta_{ij})=\frac{1}{1+\left(\frac{\Delta\theta_{ij}}{\pi}\right)^{1/T} \frac{N \bar{k}}{(1-1/T)k_i k_j}}.
\end{equation}
Therefore, in this formulation we sample the nodes' angular coordinates and expected degrees, and connect each pair of nodes $i, j$ according to the probability in~(\ref{eq:p_s1_alt}). We note that the distribution of the $\kappa$ is not explicitly fixed here, but implicitly imposed by the distribution of the $k$. For instance, if we sample the $k_i$ from a power-law, $\tilde{\rho} (k) \propto k^{-\tilde{\gamma}}$, then the distribution of the $\kappa_i$ will also be a power law, $\rho(\kappa) \propto \kappa^{-\gamma}$, but with $\gamma=(\tilde{\gamma}-1)/T+1$. We work with this alternative formulation as we can explicitly fix in it the distribution of expected degrees. We also note that such formulation is not necessary in the cold regime, since there we have $k \propto \kappa$, instead of $k \propto \kappa^{1/T}$~\cite{Krioukov2010}. In other words, in contrast to the hot regime, in the cold regime the distribution of the $k$ is the same as the distribution of the $\kappa$.

\subsection{Limit to the configuration model and to classical random graphs} 

For $T \to \infty$ the connection probability in~(\ref{eq:p_s1_alt}) converges to
\begin{equation}
\label{eq:p_cm}
p(k_i, k_j) = \frac{1}{1+N \bar{k}/(k_i k_j)},
\end{equation} 
which is the connection probability in the configuration model (the ensemble of graphs with given expected degrees~\cite{ChungLu2002, Newman2004, Krioukov2010}). In this limit, only the nodes' expected degrees matter, while the similarity distances among the nodes are completely ignored. If we further let $k_i = \bar{k}$ for all nodes $i$, then the connection probability in~(\ref{eq:p_cm}) reduces to
\begin{equation}
\label{eq:p_rg}
p=\frac{1}{1+N/\bar{k}}.
\end{equation}
In this case, the nodes' expected degrees do not matter either (as they are all the same) and the connection probability becomes the connection probability in classical random graphs ${\cal G}_{N,p}$~\cite{SoRa51}, where each of the $N(N-1)/2$ pairs of $N$ nodes is connected with the same probability $p \approx \bar{k}/N$.
 
\subsection{dynamic-$\mathbb{S}^1$ model}

The dynamic-$\mathbb{S}^{1}$ model~\cite{Papadopoulos2019} models a sequence of network snapshots, $G_t$, $t=1,\ldots, \tau$. Each snapshot is a realization of the $\mathbb{S}^{1}$ model. Therefore, there are $N$ nodes that are assigned hidden variables $\kappa, \theta$ as in the $\mathbb{S}^{1}$ model, which remain fixed. The temperature $T$ is also fixed, while in general each snapshot $G_t$ can have a different average degree $\bar{k}_t, t=1, \ldots, \tau$. The snapshots are generated according to the following simple rules:
\begin{enumerate}
\item[(1)] at each time slot $t=1, \ldots, \tau$, snapshot $G_t$ starts with $N$ disconnected nodes;
\item[(2)] each pair of nodes $i, j$ connects with probability given by~(\ref{eq:p_s1});
\item[(3)] at time slot $t+1$, all the edges in snapshot $G_t$ are deleted and the process starts over again to generate snapshot $G_{t+1}$.
\end{enumerate}
To facilitate the analysis we assume that $\bar{k}_t = \bar{k}$,~$\forall t$, i.e., that all snapshots have the same average degree $\bar{k}$. We also assign expected degrees $k$ to nodes (instead of hidden degree variables $\kappa$) and connect each pair of nodes in each time slot according to the probability in~(\ref{eq:p_s1_alt}). In the configuration model and classical random graphs, we connect each pair of nodes in each time slot according to the probability in~(\ref{eq:p_cm}) and~(\ref{eq:p_rg}), respectively.

We note that the snapshots generated by the dynamic-$\mathbb{S}^1$ are conditionally independent given the nodes' hidden variables, but \emph{not} independent. In other words, even though each snapshot $G_t$ is constructed anew, there are correlations among the snapshots induced by the nodes' effective distances $\chi_{ij}$. In particular, nodes at smaller effective distances have higher chances of being connected in each snapshot.  Such correlations also exist in the configuration model, while there are no correlations among the snapshots in classical random graphs. We also note that since the snapshots in the dynamic-$\mathbb{S}^1$ model are constructed anew with the $\mathbb{S}^1$ model, we refer to the dynamical properties of the dynamic-$\mathbb{S}^1$ as dynamical properties of RHGs.

In Ref.~\cite{Papadopoulos2019} the contact and intercontact distributions in the dynamic-$\mathbb{S}^1$ have been analyzed in the cold regime. The contact distribution is the distribution of the number of consecutive time slots that a pair of nodes is in contact, i.e., connected. The intercontact distribution is the distribution of the number of consecutive time slots separating two contacts between the same pair of nodes. Next, we analyze these two distributions in the hot regime. We begin with the contact distribution. 
\section{Contact distribution}
\label{sec:contact_dist}

Consider the probability to observe a sequence of exactly $t$ consecutive slots, where two nodes $i$ and $j$ with expected degrees $k_i$ and $k_j$ and angular distance $\Delta \theta_{ij}$ are connected, $t=1, 2, \ldots, \tau-2$. This probability, denoted by $r_\textnormal{c}(t ; k_i, k_j, \Delta\theta_{ij})$, is the percentage of observation time $\tau$ where we observe a slot where these two nodes are not connected, followed by $t$ slots where they are connected, followed by a slot where they are again not connected.  

For each duration $t$, there are $\tau-t-1$ possibilities where this duration can be realized. For instance, if $t=2$ the two nodes can be disconnected in slot $s-1$, connected in slots $s$ and $s+1$, and disconnected in slot $s+2$, where $s = 2, \ldots, \tau-2$. Therefore, the percentage of observation time where a duration of $t$ slots can be realized is 
\begin{equation}
\label{eq:tau}
g_\tau(t) \equiv \frac{\tau-t-1}{\tau}.
\end{equation}
Clearly, for any finite $t$, $g_\tau(t) \to 1$ for $\tau \to \infty$. Since two nodes $i$ and $j$ are connected in each slot with probability $p(k_i, k_j, \Delta \theta_{ij})$ given by~(\ref{eq:p_s1_alt}), we can write
\begin{align}
\label{eq:p_c}
\nonumber r_\textnormal{c}(t ; k_i, k_j, \Delta\theta_{ij})&= g_\tau(t) p(k_i, k_j, \Delta \theta_{ij})^{t}\\
&\times \left[1-p(k_i, k_j, \Delta \theta_{ij})\right]^2.
\end{align}
We note that for simplicity we ignore the cases where the first (last) of the slots that two nodes can be connected starts (ends) at the beginning (end) of the observation period $\tau$. To account for these two possibilities one needs to add the extra term $(2/\tau) p(k_i, k_j, \Delta \theta_{ij})^{t} \left[1-p(k_i, k_j, \Delta \theta_{ij})\right]$ on the right hand side of~(\ref{eq:p_c}), which vanishes as $\tau$ increases.

The contact distribution, $P_{\textnormal{c}}(t)$, gives the probability that two nodes are connected for exactly $t$ consecutive slots, given that $t \geq 1$
\begin{equation}
\label{eq:p_c_norm}
P_{\textnormal{c}}(t) =\frac{r_\textnormal{c}(t)}{\sum_{j=1}^{\tau-2} r_\textnormal{c}(j)}.
\end{equation}
In the above relation, $r_\textnormal{c}(t)$ is obtained by removing the condition on $k_i, k_j$ and $\Delta \theta_{ij}$ from~(\ref{eq:p_c}),
\begin{align}
\nonumber r_\textnormal{c}(t) &= \int \int \int r_\textnormal{c}(t ; k, k', \Delta\theta)\\
&\times  \rho(k) \rho(k')
 f(\Delta \theta)  \mathrm{d} k \mathrm{d} k' \mathrm{d} \Delta\theta,
\end{align}
where $\rho(k)$ denotes the PDF of $k$, while $f(\Delta \theta)=1/\pi$ is the PDF of $\Delta \theta$.  We note that since $P_{\textnormal{c}}(t) \propto r_\textnormal{c}(t)$ we sometimes refer to $r_\textnormal{c}(t)$ as ``contact distribution".

Removing the condition on $\Delta\theta_{ij}$ from~(\ref{eq:p_c}), gives
\begin{align}
\label{eq:full_contact_integral} 
\nonumber r_\textnormal{c}(t ; k_i, k_j ) &=  \frac{1}{\pi} \int \limits_0^\pi  r_\textnormal{c}(t ; k_i, k_j, \Delta\theta) \mathrm{d} \Delta\theta\\
&= g_\tau(t)\frac{T}{T+2} \frac{N^2 \bar{k}^2 }{(1-1/T)^2 k_i^2 k_j^2}\\
\nonumber &\times {}_2 F_{1}\left[t+2, T+2; T+3; -\frac{N \bar{k} }{(1-1/T) k_i k_j}\right],
\end{align}
where ${}_2 F_1[a, b; c; z]$ is the Gauss hypergeometric function~\cite{special_functions_book}. To obtain $r_\textnormal{c}(t)$ we need to remove the condition on $k_i$ and $k_j$ from (\ref{eq:full_contact_integral}),
\begin{equation}
\label{eq:int_exact}
r_\textnormal{c}(t) = \int \int r_\textnormal{c}(t; k, k') \rho(k) \rho(k') \mathrm{d} k \mathrm{d} k'.
\end{equation}
However,~(\ref{eq:full_contact_integral}) is not in closed-form, which renders analytic calculations of the integral in~(\ref{eq:int_exact}) non-obvious, and dependent on the form of $\rho(k)$. An exception is the simplest case, $\rho(k)=\delta(k-k')$, where $\delta$ is the dirac delta function. In that case, $r_\textnormal{c}(t)=r_\textnormal{c}(t ; k', k')$. Yet, the shape of $r_\textnormal{c}(t ; k', k')$ is still not apparent, and dependent on the values of $N, \bar{k}$ and $T$.

Given the above considerations, we do not analyze~(\ref{eq:int_exact}) for any network size $N$. Instead, we analyze~(\ref{eq:int_exact}) in the thermodynamic limit ($N \to \infty)$. Analytic results at this limit can approximate $r_\textnormal{c}(t)$ in sufficiently sparse networks. Next, we show that for $N \to \infty$ a generic analysis of $r_\textnormal{c}(t)$ that does not depend on the form of $\rho(k)$ is possible as long as $\rho(k)$ has a finite $T^{\textnormal{th}}$ moment. In the analysis that follows, symbol ``$\approx$" means \emph{approximately equal}. Further, symbol ``$\propto$" means \emph{proportional to}, i.e., $f(t) \propto h(t)$ means $f(t) = c h(t)$, where $c$ is a constant, $0 < c < \infty$.

\subsection{Expected degree distributions with finite $T^{\textnormal{th}}$ moment}
\label{sec:generic_contact}

Ignoring the prefactor $g_\tau(t)$, the dominant term in the Taylor series expansion of~(\ref{eq:full_contact_integral}) for $N \to \infty$ is
\begin{widetext}
\begin{equation}
\label{eq:dominants}
\frac{T\Gamma{(T+2)(1-1/T)^T}}{N^T\bar{k}^T} \frac{\Gamma{(t-T)}}{\Gamma{(t+2)}}k_i^T k_j^T+\frac{T}{T-t} \frac{(1-1/T)^t}{N^t \bar{k}^t}k_i^t k_j^t,
\end{equation}
\end{widetext}
where $\Gamma$ is the gamma function. We can identify the following two regimes. If $t > T$, then the first term in~(\ref{eq:dominants}) dominates for $N \to \infty$, and we can write
\begin{align}
\label{eq:l1}
\nonumber \lim_{N \to \infty} N^{T} r_\textnormal{c}(t ; k_i, k_j ) & = g_\tau(t)\frac{T\Gamma{(T+2)(1-1/T)^T}}{\bar{k}^T}\\
&\times\frac{\Gamma{(t-T)}}{\Gamma{(t+2)}} k_i^T k_j^T.
\end{align}
On the other hand, if $t < T$, then the second term in~(\ref{eq:dominants}) dominates, and we have
\begin{align}
\label{eq:l2}
 \lim_{N \to \infty} N^t r_\textnormal{c}(t ; k_i, k_j ) = g_\tau(t)\frac{T}{T-t} \frac{(1-1/T)^t}{\bar{k}^t}k_i^t k_j^t.
\end{align}
The corresponding expression for $t=T$ can be obtained by computing the limit of (\ref{eq:dominants}) for $t \to T$. Therefore, for $t > T$  we can write
\begin{align}
\label{eq:l1_uncon_tmp}
\nonumber \lim_{N \to \infty} N^{T} r_\textnormal{c}(t) & =\lim_{N \to \infty} \int \int  N^{T} r_\textnormal{c}(t; k, k') \rho(k) \rho(k') \mathrm{d} k \mathrm{d} k'\\
\nonumber & = \int \int \lim_{N \to \infty} N^{T} r_\textnormal{c}(t; k, k') \rho(k) \rho(k') \mathrm{d} k \mathrm{d} k'\\
\nonumber & = g_\tau(t) \frac{T\Gamma{(T+2)} (1-1/T)^T}{\bar{k}^T}\\
 & \times \frac{\Gamma{(t-T)}}{\Gamma{(t+2)}}(\overline{k^T})^2,
\end{align}
while for $t < T$ we have
\begin{align}
\label{eq:l2_uncon_tmp}
\nonumber \lim_{N \to \infty} N^{t} r_\textnormal{c}(t) & =\lim_{N \to \infty} \int \int  N^{t} r_\textnormal{c}(t; k, k') \rho(k) \rho(k') \mathrm{d} k \mathrm{d} k'\\
\nonumber & = \int \int \lim_{N \to \infty} N^{t} r_\textnormal{c}(t; k, k') \rho(k) \rho(k') \mathrm{d} k \mathrm{d} k'\\
 & =  g_\tau(t)\frac{T}{T-t} \frac{(1-1/T)^t}{\bar{k}^t} (\overline{k^t})^2.
\end{align}
In Appendix~\ref{sec:app1} we show that we can indeed exchange the order of the limit with the integral in~(\ref{eq:l1_uncon_tmp}) and (\ref{eq:l2_uncon_tmp}) if the distribution of expected degrees $\rho(k)$ has a finite $T^{\textnormal{th}}$ moment, i.e., if the moment
\begin{equation}
\overline{k^T} = \int k^T \rho(k) \mathrm{d}k
\end{equation}
is finite.

Eqs.~(\ref{eq:l1_uncon_tmp}) and (\ref{eq:l2_uncon_tmp}) mean that for sufficiently large $N$ we can write
\begin{align}
\label{eq:l1_uncon}
\nonumber r_\textnormal{c}(t) & \approx g_\tau(t) \frac{T\Gamma{(T+2)} (1-1/T)^T (\overline{k^T})^2}{N^T \bar{k}^T} \frac{\Gamma{(t-T)}}{\Gamma{(t+2)}}\\
& \propto g_\tau(t) \frac{\Gamma{(t-T)}}{\Gamma{(t+2)}}
\end{align}
for $t > T$, while for $t < T$
\begin{align}
\label{eq:l2_uncon}
r_\textnormal{c}(t) \approx  g_\tau(t)\frac{T}{T-t} \frac{(1-1/T)^t (\overline{k^t})^2}{N^t \bar{k}^t}.
\end{align}
For $t \ll \tau$, $g_\tau(t) \approx 1$, while $\Gamma{(t-T)}/\Gamma{(t +2)} \approx t^{-(2+T)}$ for $t \gg T$. Therefore, from~(\ref{eq:l1_uncon}), the contact distribution decays as a power law with exponent $2+T$,
\begin{equation}
\label{eq:pl}
P_{\textnormal{c}}(t) \propto  \frac{1}{t^{2+T}},
\end{equation}
for $T \ll t \ll \tau$. This behavior is similar as in the cold regime, where $P_{\textnormal{c}}(t)$ also decays as a power law with exponent $2+T$, irrespective of the form of $\rho(k)$~\cite{Papadopoulos2019}.  However, in the cold regime the exponent $2+T$ is between two and three, while here it is greater than three. Further, in the cold regime this result holds as long as $\bar{k}$ is finite, while here it holds if $\overline{k^T}$ is finite.

\begin{figure*}

\includegraphics[width=2.3in]{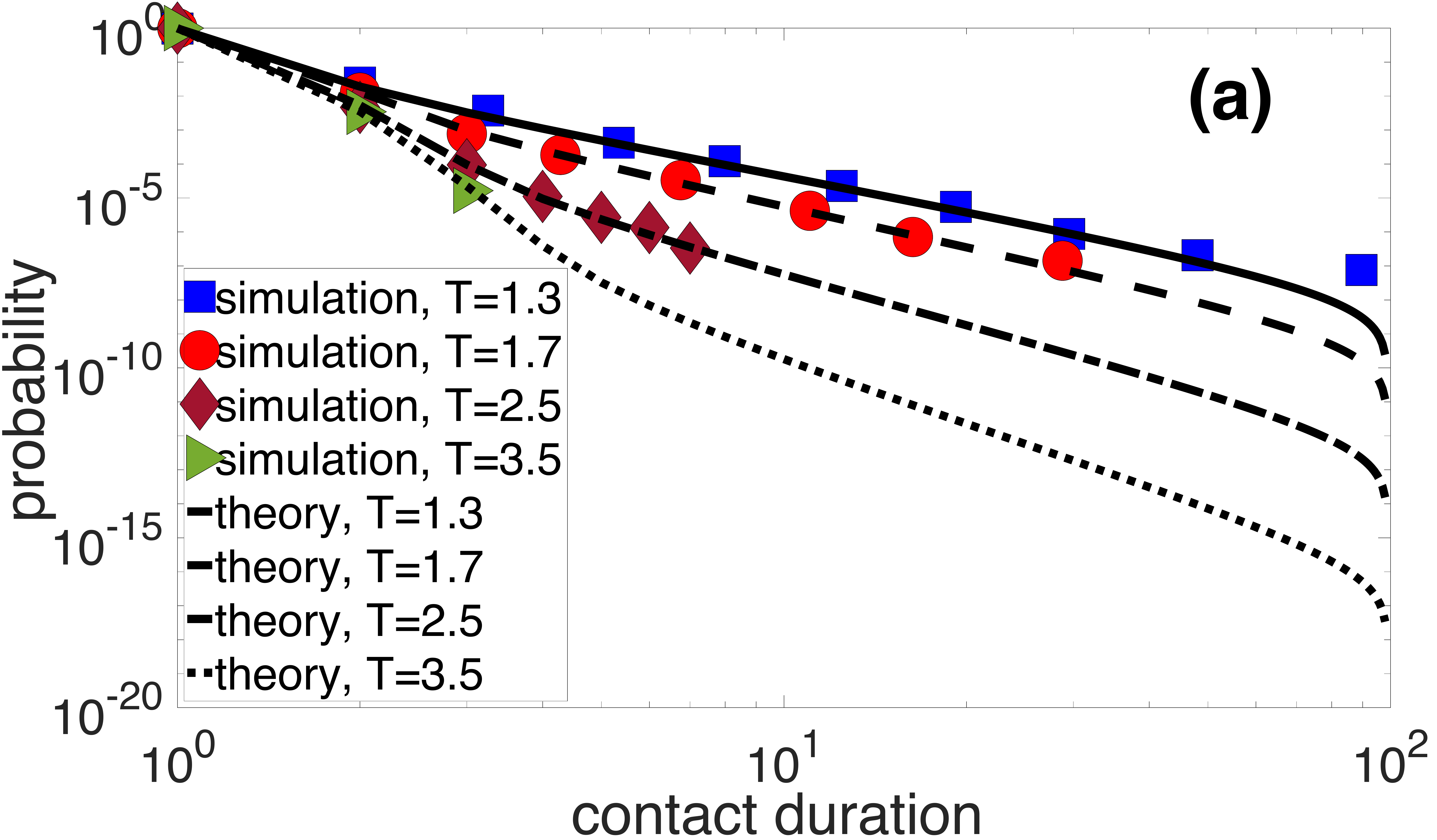}
\includegraphics[width=2.3in]{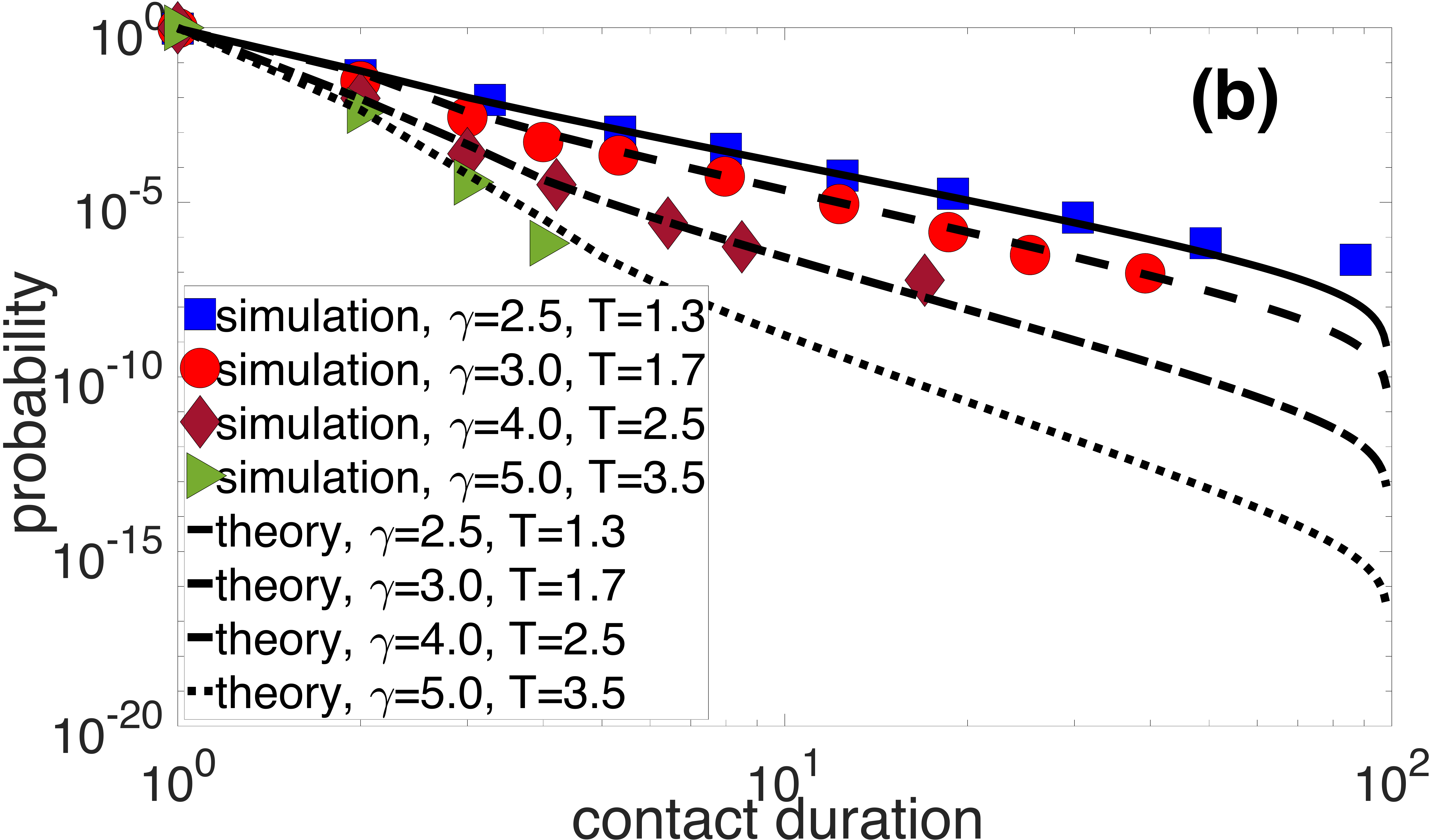}
\includegraphics[width=2.3in]{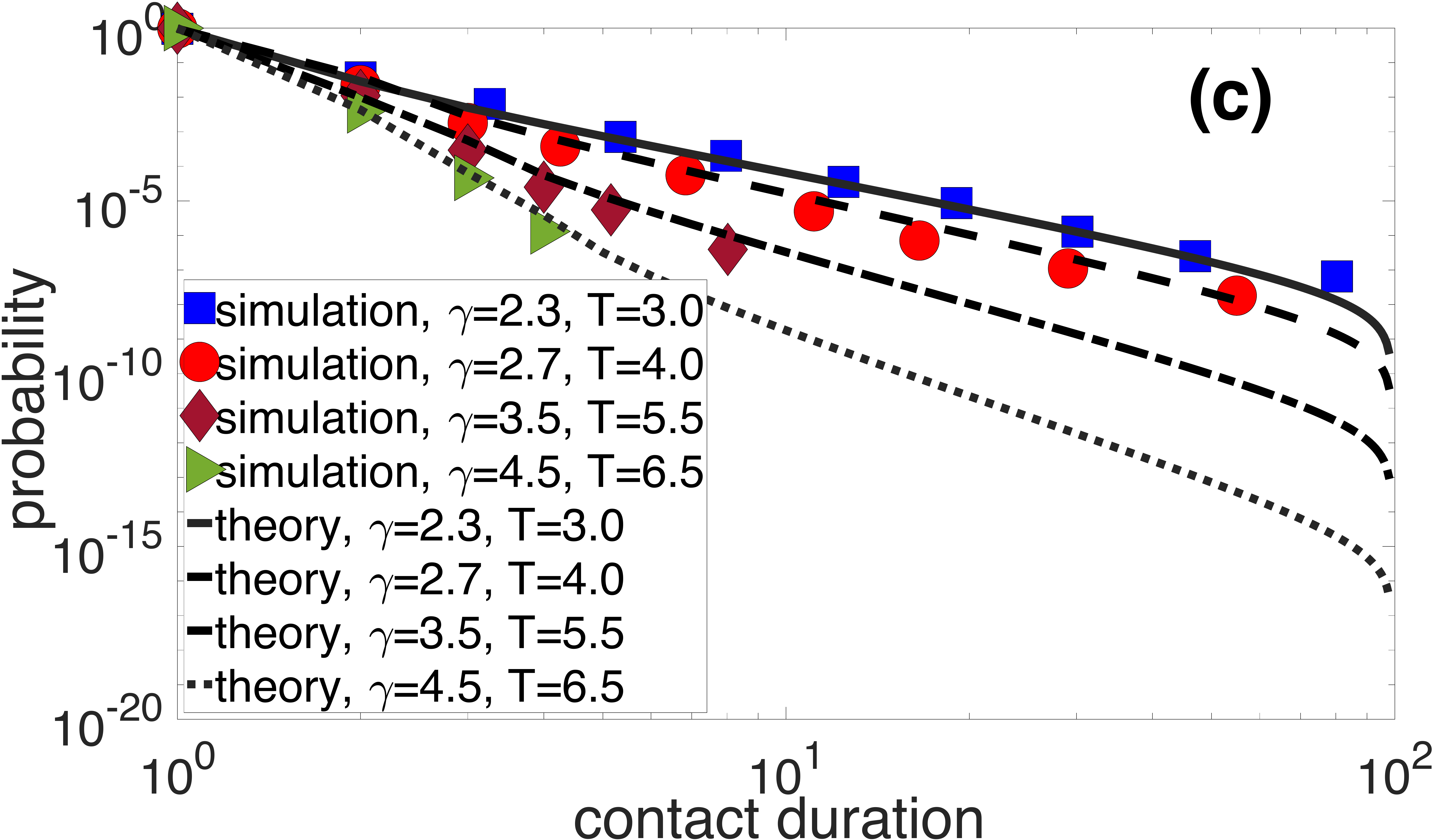}
\caption{Contact distribution in simulated networks with the dynamic-$\mathbb{S}^1$ model~vs.~theoretical predictions given by~(\ref{eq:p_c_norm}). The number of nodes is $N=2000$, the average node degree is $\bar{k}=6$, the network temperature $T$ is shown in the legends and $\tau=100$. The theoretical predictions are given by the solid and dashed/dotted lines. In~(a), the distribution of expected degrees is $\rho(k)=\delta(k-\bar{k})$, and $r_\textnormal{c}(t)$ is given by~(\ref{eq:ex1}) for contact durations $t < T$, and by~(\ref{eq:l1_uncon}) for $t > T$. In (b), the distribution of expected degrees is power law, $\rho(k) \propto k^{-\gamma}$, with $\gamma > T+1$ (shown in the legend); in this case the $T^{\textnormal{th}}$ moment of $\rho(k)$ is finite, and $r_\textnormal{c}(t)$ is given by~(\ref{eq:ex3a}) for $t < T$, and by~(\ref{eq:l1_uncon}) for $t > T$. In (c), the distribution of expected degrees is power law, $\rho(k) \propto k^{-\gamma}$, with $\gamma < T+1$; in this case the $T^{\textnormal{th}}$ moment of $\rho(k)$ is not finite, and $r_\textnormal{c}(t)$ is given by~(\ref{eq:ex3a}) for $t < \gamma-1$, and by~(\ref{eq:l3_pl}) for $t > \gamma-1$. The simulation results are averages over five runs and empirical distributions with durations $t \geq 10$ have been binned logarithmically. All axes are in logarithmic scale.
\label{fig:validation1}}
\end{figure*}

For $t < T$, the form of $P_{\textnormal{c}}(t)$ depends on the form of $\rho(k)$, as it depends on the moments $\overline{k^t}$, see~(\ref{eq:l2_uncon}) . In the simplest case, $\rho(k)=\delta(k-\bar{k})$, and~(\ref{eq:l2_uncon}) gives
\begin{equation}
\label{eq:ex1}
r_\textnormal{c}(t) \approx  g_\tau(t)\frac{T}{T-t} \eta_N^{-t},
\end{equation}
where
\begin{equation}
\label{eq:eta_N}
\eta_N \equiv \frac{N}{(1-1/T)\bar{k}}.
\end{equation}
As another example, if $\rho(k)$ is the exponential distribution, $\rho(k) =(1/\bar{k}) e^{-k/\bar{k}}$, then $\overline{k^t}= \bar{k}^t \Gamma{(t+1)}$, and~(\ref{eq:l2_uncon}) yields
\begin{equation}
\label{eq:ex2}
r_\textnormal{c}(t) \approx  g_\tau(t)\frac{T \Gamma{(t+1)}^2}{T-t}\eta_N^{-t}.
\end{equation}
Finally, if $\rho(k)$ is a power law, $\rho(k)=k_0^{\gamma-1} (\gamma-1) k^{-\gamma}$, $k \geq k_0$, $\gamma > 2$, then its $T^{\textnormal{th}}$ moment is finite only if $\gamma > T+1$. Noting that $\bar{k}=k_0(\gamma-1)/(\gamma-2)$, in this case for $t < T$ we have
\begin{equation}
\label{eq:pl_moments}
\overline{k^t}= \bar{k}^t \left(\frac{\gamma-2}{\gamma-1} \right)^t \frac{\gamma-1}{\gamma-t-1},
\end{equation}
and~(\ref{eq:l2_uncon}) yields
\begin{equation}
\label{eq:ex3a}
r_\textnormal{c}(t) \approx g_\tau(t)\frac{T (\gamma-1)^{2-2t} (\gamma-2)^{2t}}{(T-t)(\gamma-t-1)^2} \eta_N^{-t} .
\end{equation}
We observe that all three Eqs.~(\ref{eq:ex1}),~(\ref{eq:ex2}), and~(\ref{eq:ex3a}), contain the exponentially-decaying term $\eta_N^{-t}$, which is multiplied by some additional $t$-dependent factors. 
In~Fig.~\ref{fig:validation1}(a,b) we validate the above analysis with simulations.

\subsection{Expected degree distributions with non-finite $T^{\textnormal{th}}$ moment}
\label{sec:contact_non_finite}

If $\rho(k)$ does not have a finite $T^{\textnormal{th}}$ moment, then we cannot exchange the order of the limit with the integral in~(\ref{eq:l1_uncon_tmp}) (Appendix~\ref{sec:app1}). In this case, to analyze $r_\textnormal{c}(t)$ for $N \to \infty$, we have to (attempt to) compute the integral in~(\ref{eq:int_exact}) for the specific $\rho(k)$ we are interested in, and then perform its Taylor series expansion for $N \to \infty$. Below, we perform this analysis for the case of power-law distributed expected degrees, $\rho(k)=k_0^{\gamma-1} (\gamma-1) k^{-\gamma}$, $k \geq k_0$, $\gamma > 2$. 

Performing the integration in~(\ref{eq:int_exact}) when $\rho(k)$ is the above power law, yields the following long but exact expression, which holds for any network size $N$
\begin{widetext}
\begin{align}
\label{eq:pl_exact}
\nonumber r_\textnormal{c}(t) &=g_\tau(t) \frac{N^2 \bar{k}^2 T^3 (\gamma-1)^2 }{k_0^4 (T-1)^2 (1+T-\gamma)^2} \Bigg\{\frac{\Gamma{(T+2)}}{\Gamma{(T+3)}} {}_2 F_{1}\left[t+2, T+2; T+3; -\frac{N \bar{k}}{(1-1/T) k_0^2}\right]\\
 \nonumber &- \frac{\Gamma{(\gamma+1)}}{\Gamma{(\gamma+2)}} {}_2 F_{1}\left[t+2, \gamma+1; \gamma+2; -\frac{N \bar{k}}{(1-1/T) k_0^2}\right]\\
 &+ \frac{(1+T-\gamma)}{(\gamma+1)^2} {}_3 F_{2}\left[t+2, \gamma+1, \gamma+1; \gamma+2, \gamma+2;  -\frac{N \bar{k}}{(1-1/T) k_0^2}\right]\Bigg\},
\end{align}
\end{widetext}
where  ${}_3 F_{2} [a_1, a_2, a_3; b_1, b_2; z]$ is the generalized hypergeometric function~\cite{special_functions_book}. After performing the Taylor series expansion of~(\ref{eq:pl_exact}) for $N \to \infty$, we can identify the following three regimes:
\begin{itemize}
\item[(R1)] for $t < T$ and $t < \gamma-1$, we get the limit in~(\ref{eq:l2_uncon_tmp}) and thus the approximation in~(\ref{eq:ex3a});
\item[(R2)] for $t > T$ and $\gamma > T+1$, we get the limit in~(\ref{eq:l1_uncon_tmp}) and thus the approximation~(\ref{eq:l1_uncon});
\item[(R3)] for $\gamma < T +1$ and $t > \gamma-1$, we get the following limit:
\begin{align}
\label{eq:l3_pl_tmp}
\nonumber \lim_{N \to \infty}  \frac{\eta_N ^{\gamma-1}}{q_N} r_\textnormal{c}(t) &= g_\tau(t) \frac{T \Gamma{(\gamma+1)} (\gamma-1)^{2-2(\gamma-1)}}{T+1-\gamma}\\
 & \times  (\gamma-2)^{2(\gamma-1)} \frac{\Gamma{(t+1-\gamma)}}{\Gamma{(t+2)}},
\end{align}
where
\begin{equation}
\label{eq:q_N}
q_N \equiv \ln{\frac{\eta_N (\gamma-1)^2}{(\gamma-2)^2}}.
\end{equation}
\end{itemize}
Eq.~(\ref{eq:l3_pl_tmp}) means that for sufficiently large $N$ we can write
\begin{align}
\label{eq:l3_pl}
\nonumber r_\textnormal{c}(t) & \approx g_\tau(t) \frac{T \Gamma{(\gamma+1)} (\gamma-1)^{2-2(\gamma-1)} (\gamma-2)^{2(\gamma-1)} }{T+1-\gamma}\\
\nonumber & \times \eta_N^{-(\gamma-1)} q_N  \frac{\Gamma{(t+1-\gamma)}}{\Gamma{(t+2)}}\\
&  \propto g_\tau(t) \frac{\Gamma{(t+1-\gamma)}}{\Gamma{(t+2)}}.
\end{align}
We note that for $\gamma > T+1$, the $T^{\textnormal{th}}$ moment of $\rho(k)$ is finite, and regimes (R1) and (R2), found also by the analysis of Sec.~\ref{sec:generic_contact}, apply. On the other hand, in regime (R3), $\gamma < T+1$, and thus the $T^{\textnormal{th}}$ moment of $\rho(k)$ is not finite. This regime did not emerge in the analysis of Sec.~\ref{sec:generic_contact}. We also note that (R1) applies if $\gamma < T+1$ and $t < \gamma-1$.

\begin{figure}
\includegraphics[width=3in]{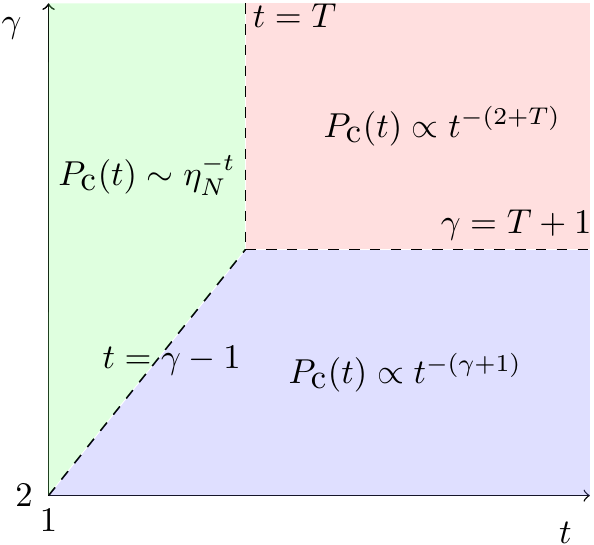}
\caption{Regimes of the contact distribution $P_\textnormal{c}(t)$ for expected node degrees distributed as a power-law with exponent $\gamma > 2$. The vertical and horizontal dashed lines correspond respectively to $t=T$ and $\gamma=T+1$. The diagonal dashed line corresponds to $t=\gamma-1$. The green, red and blue shaded regions indicate respectively the three regimes given by~(\ref{eq:ex3a}),~(\ref{eq:l1_uncon}) and~(\ref{eq:l3_pl}). In the power-law regimes (red and blue shaded regions) the exponents $2+T$ and $\gamma+1$ are always greater than three. 
\label{fig:pl_regimes}}
\end{figure}

For $t \ll \tau$, $g_\tau(t) \approx 1$, while $\Gamma{(t+1-\gamma)}/\Gamma{(t+2)} \approx t^{-(\gamma+1)}$ for $t \gg \gamma$. Therefore, from~(\ref{eq:l3_pl}),  in~(R3) the contact distribution decays as a power law with exponent $\gamma+1$,
\begin{equation}
\label{eq:pl2}
P_{\textnormal{c}}(t) \propto  \frac{1}{t^{\gamma+1}},
\end{equation}
for $\gamma \ll t \ll \tau$. Since $\gamma > 2$, the exponent of this power law is again larger than three, while there is no analogous behavior in the cold regime~\cite{Papadopoulos2019}. Fig.~\ref{fig:validation1}(c) validates regime (R3) with simulations, while Fig.~\ref{fig:pl_regimes} provides a schematic overview of the three contact distribution regimes.

We note that (\ref{eq:pl_exact}) does not hold for $\gamma=T+1$. The corresponding relation for $\gamma=T+1$ can be obtained by letting $T=\gamma-1$ in~(\ref{eq:full_contact_integral}), and then performing the integration in~(\ref{eq:int_exact}). In this case, for $N \to \infty$ we get regime (R1) for $t < \gamma-1$, while for $t > \gamma-1$ we get
\begin{align}
\label{eq:l3_plgT}
\nonumber  \lim_{N \to \infty} \frac{\eta_N ^{\gamma-1}}{q_N^2}  r_\textnormal{c}(t) & = g_\tau(t) \Gamma{(\gamma+1)} (\gamma-1)^{3-2(\gamma-1)}\\
& \times (\gamma-2)^{2(\gamma-1)} \frac{\Gamma{(t+1-\gamma)}}{2\Gamma{(t+2)}},
\end{align}
which is similar to (R3).

\subsection{Contact distribution in the configuration model}
\label{sec:contact_cm}

The contact distribution in the configuration model for large $N$ can be obtained by letting $T \to \infty$ in the corresponding finite-$T$ relations. Let us consider again the case of power-law distributed expected degrees. Since $T \to \infty$, only regimes (R1) and (R3) from Sec.~\ref{sec:contact_non_finite} are relevant. For $t < \gamma-1$, $r_\textnormal{c}(t)$ is obtained by letting $T \to \infty$ in~(\ref{eq:ex3a}),
\begin{equation}
\label{eq:ex3a_cm}
r_\textnormal{c}(t) \approx g_\tau(t)\frac{(\gamma-1)^{2-2t} (\gamma-2)^{2t}}{(\gamma-t-1)^2} \left(\frac{N}{\bar{k}}\right)^{-t}.
\end{equation}
For $t > \gamma-1$, $r_\textnormal{c}(t)$ is obtained by letting $T \to \infty$ in~(\ref{eq:l3_pl}), 
\begin{align}
\label{eq:l3_pl_cm}
\nonumber r_\textnormal{c}(t) & \approx g_\tau(t) \Gamma{(\gamma+1)} (\gamma-1)^{2-2(\gamma-1)} (\gamma-2)^{2(\gamma-1)} \\
\nonumber & \times (N/\bar{k})^{-(\gamma-1)}\ln{\frac{N (\gamma-1)^2}{\bar{k}(\gamma-2)^2}} \frac{\Gamma{(t+1-\gamma)}}{\Gamma{(t+2)}}\\
& \propto g_\tau(t)\frac{\Gamma{(t+1-\gamma)}}{\Gamma{(t+2)}}.
\end{align}
Therefore, $P_\textnormal{c}(t) \propto t^{-(\gamma+1)}$ for $\gamma \ll t \ll \tau$. We can also obtain the exact relation for $r_\textnormal{c}(t)$ for any network size $N$ by letting $T \to \infty$ in~(\ref{eq:pl_exact}),
\begin{align}
\label{eq:pl_exact_cm}
r_\textnormal{c}(t) &=g_\tau(t) \frac{N^2 \bar{k}^2 (\gamma-1)^2 }{k_0^4 (\gamma+1)^2}\\
\nonumber &\times {}_3 F_{2}\left[t+2, \gamma+1, \gamma+1; \gamma+2, \gamma+2;  -\frac{N \bar{k}}{k_0^2}\right].
\end{align}
The above relation can be also obtained directly by using in~(\ref{eq:p_c}) the connection probability in~(\ref{eq:p_cm}) (instead of the connection probability in~(\ref{eq:p_s1_alt})), and computing the integral in~(\ref{eq:int_exact}) with the power-law $\rho(k)$.

\subsection{Contact distribution in classical random graphs}
\label{sec:contact_rgs}

Finally, the contact distribution in classical random graphs for large $N$ can be obtained by letting $T \to \infty$ in~(\ref{eq:ex1}),
\begin{equation}
\label{eq:ex1_rg}
r_\textnormal{c}(t) \approx  g_\tau(t)\left(\frac{N}{\bar{k}}\right)^{-t}.
\end{equation}
We can also deduce the above expression directly from the exact relation for $r_\textnormal{c}(t)$ in classical random graphs. Specifically, in classical random graphs the connection probability is $p \approx \bar{k}/N$, while~(\ref{eq:p_c}) becomes
\begin{equation}
\label{eq:exact_er_c}
r_\textnormal{c}(t) = g_\tau(t) p^{t} (1-p)^2.
\end{equation}
The above expression is a product of the linearly-decaying term $g_\tau(t)$ and the exponentially-decaying term $p^t=(N/\bar{k})^{-t}$. For sufficiently large $N$, $(1-p)^2 \approx 1$, resulting in~(\ref{eq:ex1_rg}). Next, we turn our attention to the intercontact distribution.

\section{Intercontact distribution}
\label{sec:intercontact_dist}

To analyze the intercontact distribution we follow a similar approach as in the contact distribution. Let $r_\textnormal{ic}(t; k_i, k_j, \Delta\theta_{ij})$ be the probability to observe a sequence of exactly $t$ consecutive slots, where two nodes  $i$ and $j$ with expected degrees $k_i$ and $k_j$ and angular distance $\Delta \theta_{ij}$ are not connected, $t=1, 2, \ldots, \tau-2$. This probability is the percentage of observation time $\tau$ where we observe a slot where these two nodes are connected, followed by $t$ slots where they are not connected, followed by a slot where they are again connected. We can write
\begin{align}
\label{eq:p_ic}
\nonumber r_\textnormal{ic}(t ; k_i, k_j, \Delta\theta_{ij})&= g_\tau(t) p(k_i, k_j, \Delta \theta_{ij})^2\\
&\times [1-p(k_i, k_j, \Delta \theta_{ij})]^t.
\end{align}
The intercontact distribution, $P_{\textnormal{ic}}(t)$, gives the probability that two nodes are disconnected for exactly $t$ consecutive slots, given that $t \geq 1$
\begin{equation}
\label{eq:p_ic_norm}
P_{\textnormal{ic}}(t) =\frac{r_\textnormal{ic}(t)}{\sum_{j=1}^{\tau-2} r_\textnormal{ic}(j)}.
\end{equation}
In the last relation, $r_\textnormal{ic}(t)$ is obtained by removing the condition on $k_i, k_j$ and $\Delta \theta_{ij}$ from~(\ref{eq:p_ic}),
\begin{align}
\nonumber r_\textnormal{ic}(t) &= \int \int \int r_\textnormal{ic}(t ; k, k', \Delta\theta)\\
 &\times  \rho(k) \rho(k')
 f(\Delta \theta)  \mathrm{d} k \mathrm{d} k' \mathrm{d} \Delta\theta.
\end{align}
Since $P_{\textnormal{ic}}(t) \propto r_\textnormal{ic}(t)$ we sometimes refer to $r_\textnormal{ic}(t)$ as ``intercontact distribution". 

Removing the condition on $\Delta \theta_{ij}$ from~(\ref{eq:p_ic}), gives
\begin{align}
\label{eq:full_intercontact_integral}
\nonumber r_\textnormal{ic}(t ; k_i, k_j ) &=  \frac{1}{\pi} \int \limits_0^\pi  r_\textnormal{ic}(t ; k_i, k_j, \Delta\theta) \mathrm{d} \Delta\theta\\
&= g_\tau(t)\frac{T}{t+T} \frac{N^t \bar{k}^t }{(1-1/T)^t k_i^t k_j^t}\\
\nonumber &\times {}_2 F_{1}\left[t+2, t+T; t+T+1; -\frac{N \bar{k}}{(1-1/T) k_i k_j}\right].
\end{align}
To obtain $r_\textnormal{ic}(t)$ we need to remove the condition on $k_i$ and $k_j$ from~(\ref{eq:full_intercontact_integral}),
\begin{equation}
\label{eq:int_ic_exact}
r_\textnormal{ic}(t) = \int \int r_\textnormal{ic}(t; k, k') \rho(k) \rho(k') \mathrm{d} k \mathrm{d} k'.
\end{equation}
However, (\ref{eq:full_intercontact_integral}) is not in closed-form, and as in the case of the contact distribution, we analyze $r_\textnormal{ic}(t)$ for $N \to \infty$. 
Below, we show that for $N \to \infty$ a generic analysis of $r_\textnormal{ic}(t)$ that does not depend on $\rho(k)$ is possible, as long as $\rho(k)$ has a finite $T^{\textnormal{th}}$ moment if $T \in (1, 2)$, or a finite $2^{\textnormal{nd}}$ moment if $T > 2$. 

\subsection{Expected degree distributions with finite moments}
\label{sec:generic_intercontact}

Ignoring the prefactor $g_\tau(t)$, the dominant term in the Taylor series expansion of (\ref{eq:full_intercontact_integral}) for $N \to \infty$ is
\begin{widetext}
\begin{equation}
\label{eq:dominants_ic}
\frac{T \Gamma{(2-T)}(1-1/T)^T}{N^T \bar{k}^T}\frac{\Gamma{(t+T)}}{\Gamma{(t+2)}} k_i^T k_j^T+\frac{T}{T-2}\frac{(1-1/T)^2}{N^2 \bar{k}^2}k_i^2 k_j^2.
\end{equation}
\end{widetext}
Here, we can identify the following two regimes. If $T \in(1, 2)$, then the first term in~(\ref{eq:dominants_ic}) dominates for $N \to \infty$, and we can write
\begin{align}
\label{eq:l1_ic}
\nonumber \lim_{N \to \infty} N^{T} r_\textnormal{ic}(t ; k_i, k_j )  & = g_\tau(t)\frac{T \Gamma{(2-T)}(1-1/T)^T}{\bar{k}^T}\\
&\times \frac{\Gamma{(t+T)}}{\Gamma{(t+2)}} k_i^T k_j^T.
\end{align}
On the other hand, if $T > 2$, then the second term in~(\ref{eq:dominants_ic}) dominates, and we have
\begin{align}
\label{eq:l2_ic}
\lim_{N \to \infty} N^{2} r_\textnormal{ic}(t ; k_i, k_j ) = g_\tau(t)\frac{T}{T-2}\frac{(1-1/T)^2}{\bar{k}^2}k_i^2 k_j^2.
\end{align}
The corresponding expression for $T=2$ can be obtained by computing the limit of (\ref{eq:dominants_ic}) for $T \to 2$. Therefore, for $T \in (1, 2)$  we can write
\begin{align}
\label{eq:l1_uncon_ic_tmp}
\nonumber \lim_{N \to \infty} N^{T} r_\textnormal{ic}(t) & =\lim_{N \to \infty} \int \int  N^{T} r_\textnormal{ic}(t; k, k') \rho(k) \rho(k') \mathrm{d} k \mathrm{d} k'\\
\nonumber & = \int \int \lim_{N \to \infty} N^{T} r_\textnormal{ic}(t; k, k') \rho(k) \rho(k') \mathrm{d} k \mathrm{d} k'\\
\nonumber & = g_\tau(t)\frac{T \Gamma{(2-T)}(1-1/T)^T}{\bar{k}^T}\\
&\times \frac{\Gamma{(t+T)}}{\Gamma{(t+2)}} (\overline{k^T})^2,
\end{align}
while for $T > 2$ we have
\begin{align}
\label{eq:l2_uncon_ic_tmp}
\nonumber \lim_{N \to \infty} N^{2} r_\textnormal{ic}(t) & =\lim_{N \to \infty} \int \int  N^{2} r_\textnormal{ic}(t; k, k') \rho(k) \rho(k') \mathrm{d} k \mathrm{d} k'\\
\nonumber & = \int \int \lim_{N \to \infty} N^{2} r_\textnormal{ic}(t; k, k') \rho(k) \rho(k') \mathrm{d} k \mathrm{d} k'\\
 & =  g_\tau(t)\frac{T}{T-2}\frac{(1-1/T)^2}{\bar{k}^2} (\overline{k^2})^2.
\end{align}
In Appendix~\ref{sec:app2} we show that we can indeed exchange the order of the limit with the integral in~(\ref{eq:l1_uncon_ic_tmp}) if $\rho(k)$ has a finite $T^{\textnormal{th}}$ moment, and in~(\ref{eq:l2_uncon_ic_tmp}) if $\rho(k)$ has a finite  $2^{\textnormal{nd}}$ moment.

Eqs.~(\ref{eq:l1_uncon_ic_tmp}) and (\ref{eq:l2_uncon_ic_tmp}) mean that for sufficiently large $N$ we can write
\begin{align}
\label{eq:l1_uncon_ic}
\nonumber r_\textnormal{ic}(t) &\approx  g_\tau(t)\frac{T \Gamma{(2-T)}(1-1/T)^T(\overline{k^T})^2}{N^T \bar{k}^T}\frac{\Gamma{(t+T)}}{\Gamma{(t+2)}}\\
& \propto g_\tau(t) \frac{\Gamma{(t+T)}}{\Gamma{(t+2)}}
\end{align}
for $T \in (1, 2)$, while for $T > 2$
\begin{align}
\label{eq:l2_uncon_ic}
\nonumber r_\textnormal{ic}(t) &\approx  g_\tau(t)\frac{T}{T-2}\frac{(1-1/T)^2 (\overline{k^2})^2}{N^2 \bar{k}^2}\\
& \propto  g_\tau(t).
\end{align}
For $t \ll \tau$, $g_\tau(t) \approx 1$, while $\Gamma{(t+T)}/\Gamma{(t+2)} \approx t^{-(2-T)}$ for $t \gg T$. Therefore, for $T \in (1,2)$, we have from~(\ref{eq:l1_uncon_ic}) that the intercontact distribution decays as a power law with exponent $2-T$,
\begin{equation}
\label{eq:pl_ic}
P_{\textnormal{ic}}(t) \propto  \frac{1}{t^{2-T}},
\end{equation}
for $1 \ll t \ll \tau$. This behavior is again similar as in the cold regime, where $P_{\textnormal{ic}}(t)$ also decays as a power law with exponent $2-T$, irrespective of the form of $\rho(k)$~\cite{Papadopoulos2019}. However, since here $T \in (1, 2)$, the exponent $2-T$ is less than one. Power-law intercontact distributions with exponents less than one are not normalizable, in the sense that the sum in the denominator of~(\ref{eq:p_ic_norm}) tends to infinity for $\tau \to \infty$, and have not been observed in real temporal networks. By contrast, in the cold regime the exponent $2-T$ takes values between one and two, as observed in real systems~\cite{hui_paper, chaintreau_paper, Partners2011, HighSchoolData2}.  Further, in the cold regime this result holds as long as $\bar{k}$ is finite, while here it holds only if $T \in (1, 2)$ and $\overline{k^T}$ is finite.

\begin{figure*}
\includegraphics[width=2.3in]{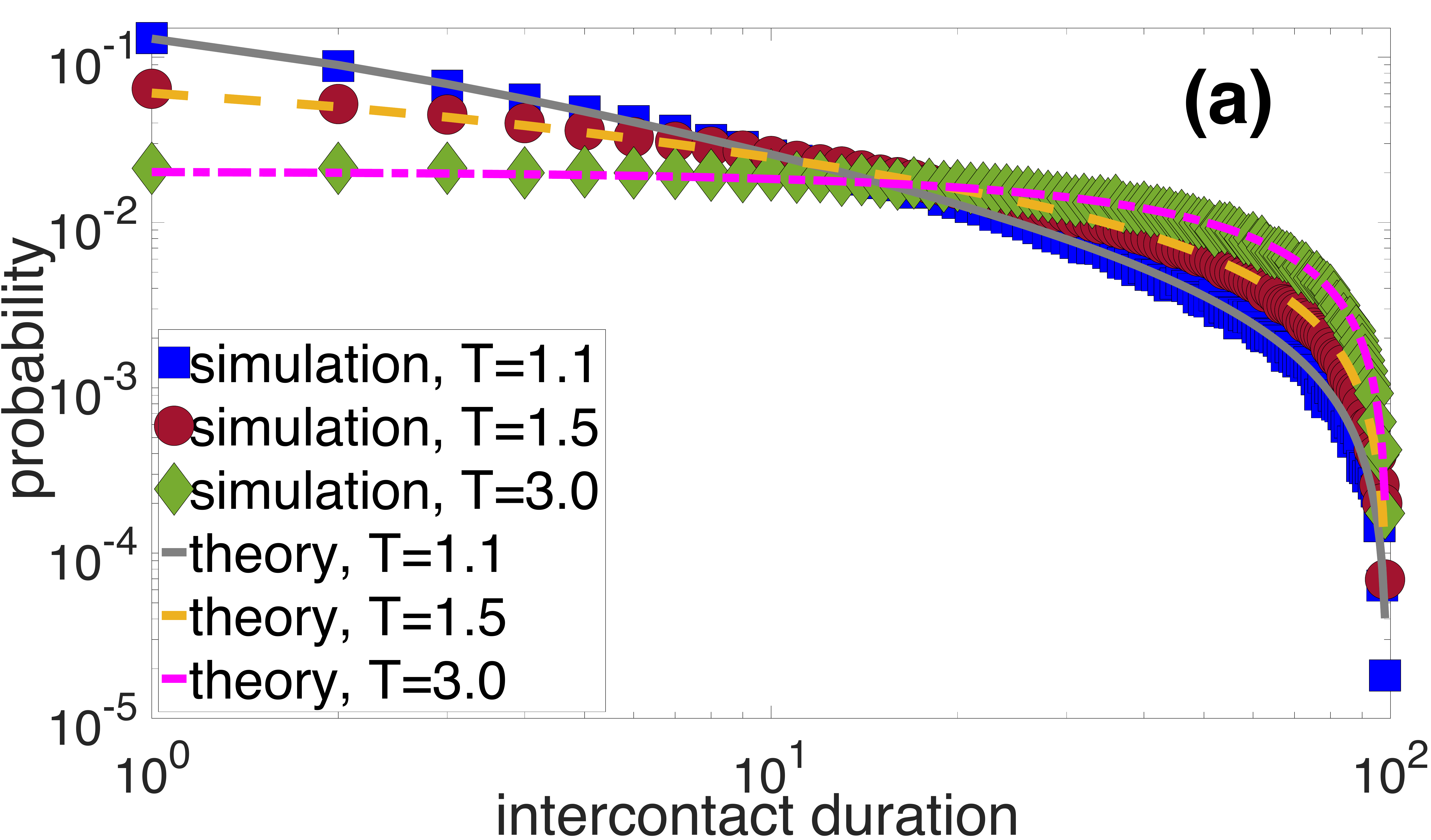}
\includegraphics[width=2.3in]{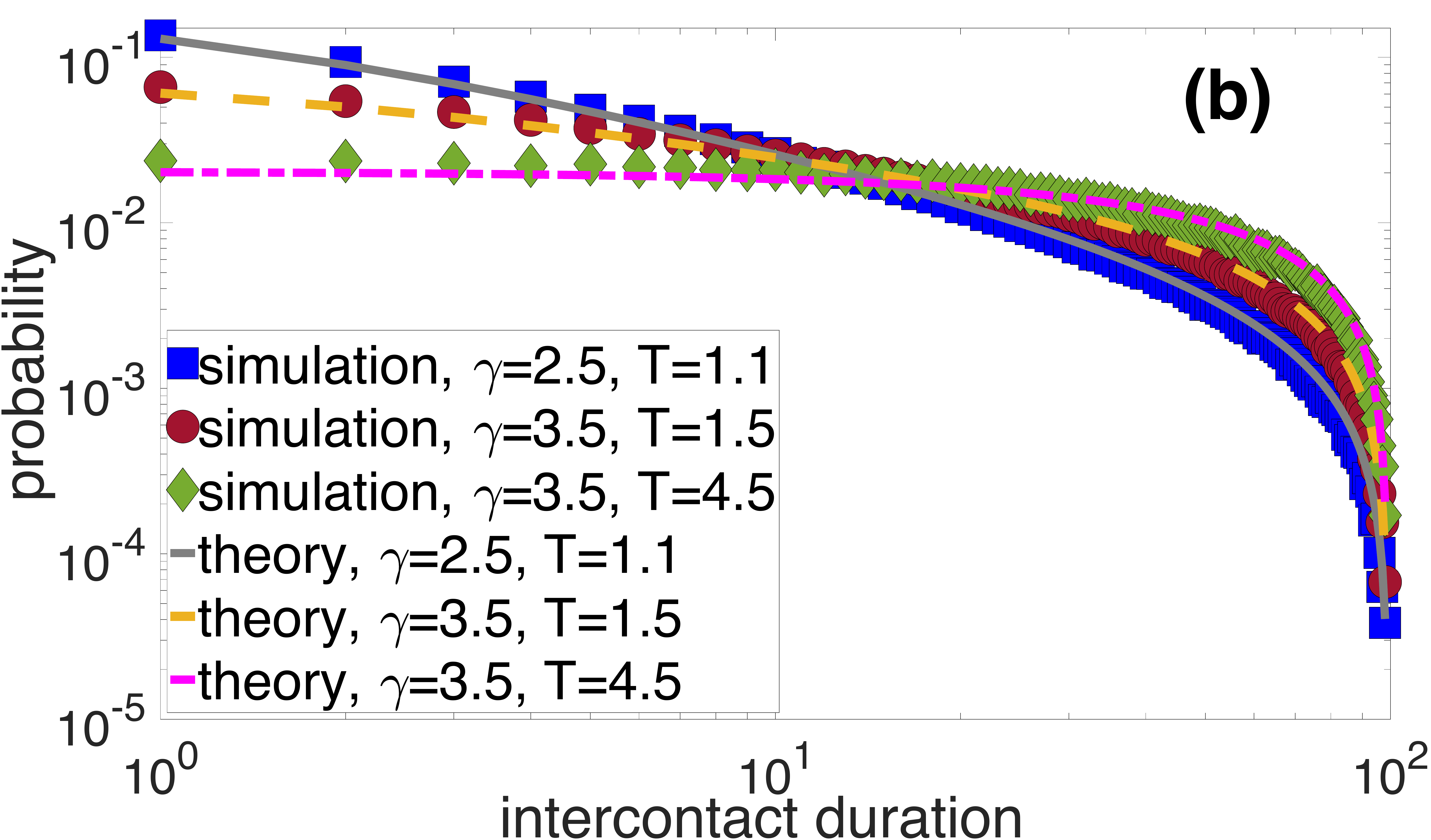}
\includegraphics[width=2.3in]{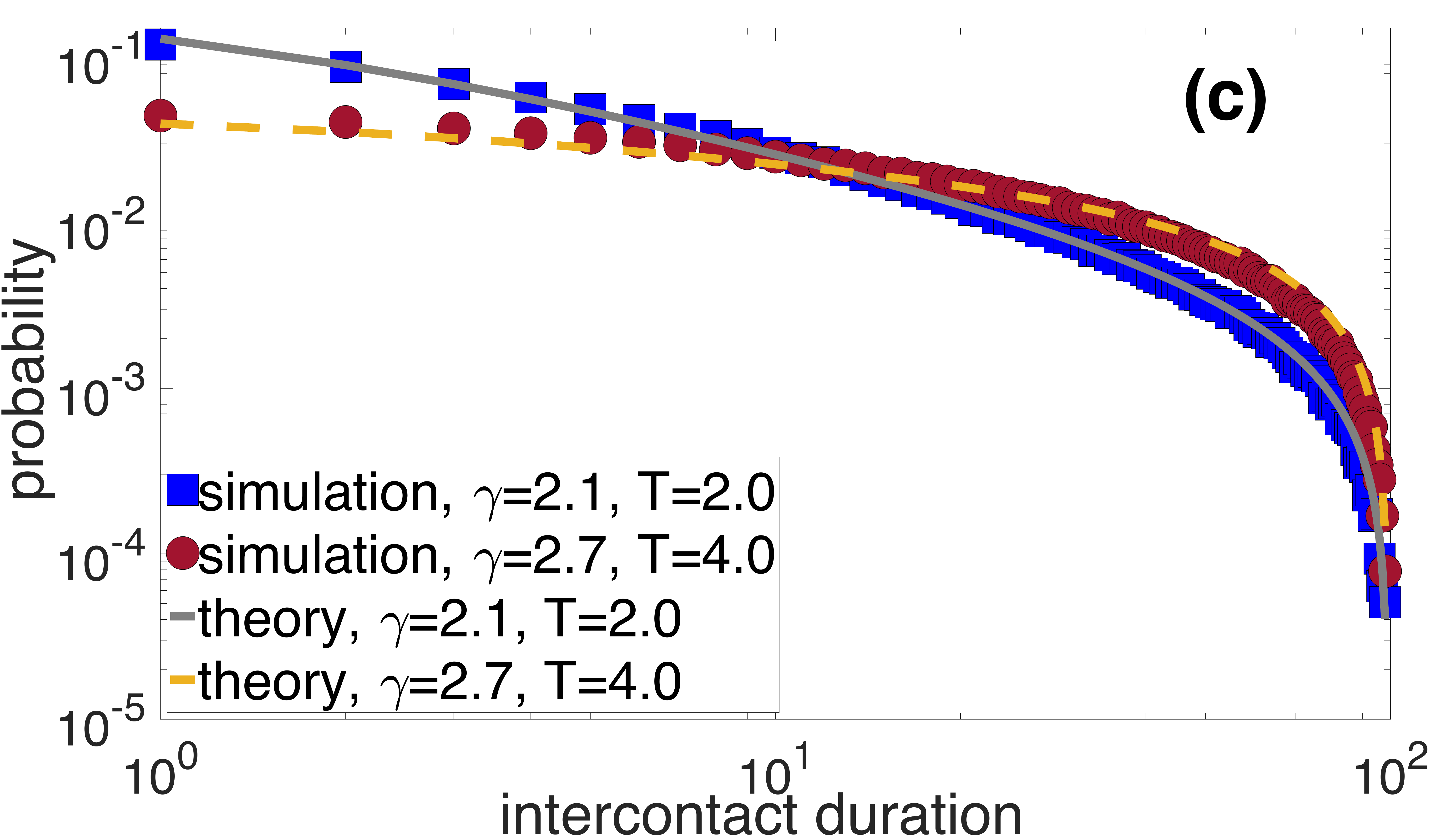}
\caption{Intercontact distribution in simulated networks with the dynamic-$\mathbb{S}^1$ model~vs.~theoretical predictions given by~(\ref{eq:p_ic_norm}). The number of nodes is $N=10000$, the average node degree is $\bar{k}=6$, the network temperature $T$ is shown in the legends and $\tau=100$. The theoretical predictions are given by the solid and dashed lines. In~(a,b), $r_\textnormal{ic}(t)$ is given by~(\ref{eq:l1_uncon_ic}) for $T \in (1, 2)$, and by~(\ref{eq:l2_uncon_ic}) for $T > 2$. In~(a), the distribution of expected degrees is $\rho(k)=\delta(k-\bar{k})$. In~(b), $\rho(k) \propto k^{-\gamma}$, with $\gamma$ as shown in the legend. In (c), $\rho(k) \propto k^{-\gamma}$ with $\gamma < T+1$ and $\gamma \in (2,3)$; in this case $r_\textnormal{ic}(t)$ is given by~(\ref{eq:l3_pl_ic}). All axes are in logarithmic scale.
\label{fig:validation2}}
\end{figure*}

For $T > 2$, we have from~(\ref{eq:l2_uncon_ic}) that $P_{\textnormal{ic}}(t)$ is proportional to the linearly-decaying term $g_\tau(t)$,
\begin{equation}
\label{eq:linear}
P_{\textnormal{ic}}(t) \propto g_\tau(t)=1-\frac{1}{\tau}(t+1).
\end{equation}
This means that in this case $P_{\textnormal{ic}}(t)$ is determined only by the boundedness of the observation interval $\tau$, displaying a linear decay with a slope that depends on $\tau$. We provide intuition on this point by explicitly showing how~(\ref{eq:linear}) emerges in classical random graphs in Sec.~\ref{sec:intercontact_rgs}. In Fig.~\ref{fig:validation2}(a,b) we validate the above analysis with simulations.

\subsection{Expected degree distributions with non-finite moments}
\label{sec:intercontact_non_finite}

If $T \in (1, 2)$ and $\overline{k^T}$ is not finite, then we cannot exchange the order of the limit with the integral in~(\ref{eq:l1_uncon_ic_tmp}) (Appendix~\ref{sec:app2}). Similarly, if $T > 2$ and $\overline{k^2}$ is not finite, we cannot exchange the order of the limit with the integral in~(\ref{eq:l2_uncon_ic_tmp}) (Appendix~\ref{sec:app2}). In these cases, to analyze $r_\textnormal{ic}(t)$ for $N \to \infty$, we have to compute the integral in~(\ref{eq:int_ic_exact}) for the specific $\rho(k)$ we are interested in, and then perform its Taylor series expansion for $N \to \infty$.  Below, we again perform this analysis for the case where $\rho(k)$ is a power-law, $\rho(k)=k_0^{\gamma-1} (\gamma-1) k^{-\gamma}$, $k \geq k_0$, $\gamma > 2$. 

Performing the integration in~(\ref{eq:int_ic_exact}) when $\rho(k)$ is the above power law, yields the following long expression that holds for any network size $N$
\begin{widetext}
\begin{align}
\label{eq:pl_exact_ic}
\nonumber r_\textnormal{ic}(t) &=g_\tau(t) \frac{N^t \bar{k}^t T^{t+1} (\gamma-1)^2 }{k_0^{2t} (T-1)^t (1+T-\gamma)^2} \Bigg\{\frac{\Gamma{(t+T)}}{\Gamma{(t+T+1)}} {}_2 F_{1}\left[t+2, t+T; t+T+1; -\frac{N \bar{k}}{(1-1/T) k_0^2}\right]\\
 \nonumber &- \frac{\Gamma{(t+\gamma-1)}}{\Gamma{(t+\gamma)}} {}_2 F_{1}\left[t+2, t+\gamma-1; t+\gamma; -\frac{N \bar{k}}{(1-1/T) k_0^2}\right]\\
 &+ \frac{(1+T-\gamma)}{(t+\gamma-1)^2} {}_3 F_{2}\left[t+2, t+\gamma-1, t+\gamma-1; t+\gamma, t+\gamma;  -\frac{N \bar{k}}{(1-1/T) k_0^2}\right]\Bigg\}.
\end{align}
\end{widetext}
After performing the Taylor series expansion of~(\ref{eq:pl_exact_ic}) for $N \to \infty$, we can identify the following three regimes:
\begin{itemize}
\item[(R1')] for $T \in (1, 2)$ and $\gamma > T+1$, we get the limit in~(\ref{eq:l1_uncon_ic_tmp}) and thus the approximation in~(\ref{eq:l1_uncon_ic});
\item[(R2')] for $ T > 2$ and $\gamma > 3$, we get the limit in~(\ref{eq:l2_uncon_ic_tmp}) and thus the approximation in~(\ref{eq:l2_uncon_ic});
\item[(R3')] for $\gamma < T +1$ and $\gamma \in (2, 3)$, we get the following limit:
\begin{align}
\label{eq:l3_pl_ic_tmp}
\nonumber  \lim_{N \to \infty}  \frac{\eta_N ^{\gamma-1}}{q_N} r_\textnormal{ic}(t) & = g_\tau(t) \Gamma{(3-\gamma)} \frac{T (\gamma-1)^{2-2(\gamma-1)}}{T+1-\gamma}\\
 & \times(\gamma-2)^{2(\gamma-1)} \frac{\Gamma{(t+\gamma-1)}}{\Gamma{(t+2)}},
\end{align}
where $\eta_N$ and $q_N$ are given by~(\ref{eq:eta_N}) and~(\ref{eq:q_N}).
\end{itemize}
Eq.~(\ref{eq:l3_pl_ic_tmp}) means that for sufficiently large $N$ we can write
\begin{align}
\label{eq:l3_pl_ic}
\nonumber r_\textnormal{ic}(t) & \approx g_\tau(t) \Gamma{(3-\gamma)}  \frac{T (\gamma-1)^{2-2(\gamma-1)} (\gamma-2)^{2(\gamma-1)}}{T+1-\gamma} \\
\nonumber & \times \eta_N ^{-(\gamma-1)} q_N \frac{\Gamma{(t+\gamma-1)}}{\Gamma{(t+2)}}\\
& \propto g_\tau(t) \frac{\Gamma{(t+\gamma-1)}}{\Gamma{(t+2)}}.
\end{align}
We note that regime (R1') corresponds to the case where $T \in (1, 2)$, while the $T^{\textnormal{th}}$ moment of $\rho(k)$ is finite; (R2') is the case where $T > 2$, while the $2^{\textnormal{nd}}$ moment of $\rho(k)$ is finite. These two regimes were also found by the analysis of Sec.~\ref{sec:generic_intercontact}. On the other hand, regime (R3') emerges in the following two cases: (i)~the case where $T \in (1, 2)$, while the $T^{\textnormal{th}}$ moment of $\rho(k)$ is not finite; and (ii)~the case where $T > 2$, while the $2^{\textnormal{nd}}$ moment of $\rho(k)$ is not finite. This regime did not emerge in the analysis of Sec.~\ref{sec:generic_intercontact}.

\begin{figure}
\includegraphics[width=3in]{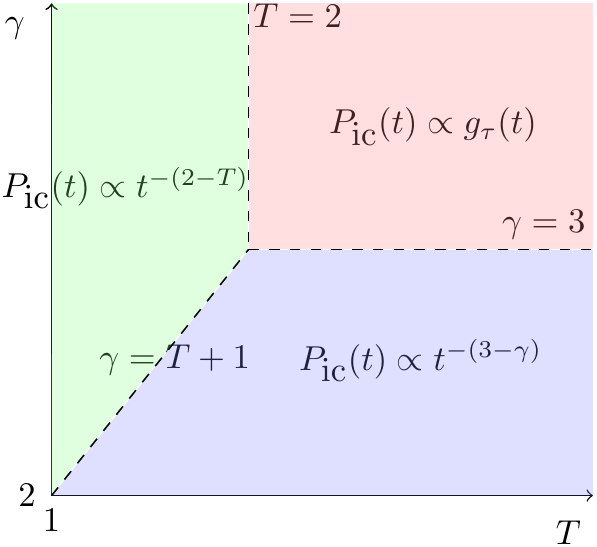}
\caption{Regimes of the intercontact distribution $P_\textnormal{ic}(t)$ for expected node degrees distributed as a power-law with exponent $\gamma > 2$. The vertical and horizontal dashed lines correspond respectively to $T=2$ and $\gamma=3$. The diagonal dashed line corresponds to $\gamma=T+1$. The green, red and blue shaded regions indicate respectively the three regimes given by~(\ref{eq:l1_uncon_ic}), (\ref{eq:l2_uncon_ic}) and~(\ref{eq:l3_pl_ic}). In the power-law regimes (green and blue shaded regions) the exponents $2-T$ and $3-\gamma$ are always less than one.
\label{fig:pl_regimes_ic}}
\end{figure}

For $t \ll \tau$, $g_\tau(t) \approx 1$, while $\Gamma{(t+\gamma-1)}/\Gamma{(t+2)} \approx t^{-(3-\gamma)}$ for $t \gg \gamma$. Therefore, from~(\ref{eq:l3_pl_ic}), in (R3') the intercontact distribution decays as a power law with exponent $3-\gamma$,
\begin{equation}
\label{eq:pl2_ic}
P_{\textnormal{ic}}(t) \propto  \frac{1}{t^{3-\gamma}},
\end{equation}
for $\gamma \ll t \ll \tau$. Since $\gamma \in (2, 3)$, the exponent of this power law is again less than one, while there is no analogous behavior in the cold regime~\cite{Papadopoulos2019}. Fig.~\ref{fig:validation2}(c) validates regime (R3') with simulations, while Fig.~\ref{fig:pl_regimes_ic} provides a schematic overview of the three intercontact distribution regimes.

Eq.~(\ref{eq:pl_exact_ic}) does not hold for $\gamma=T+1$. The corresponding relation for $\gamma=T+1$ can be obtained by letting $T=\gamma-1$ in~(\ref{eq:full_intercontact_integral}), and then performing the integration in~(\ref{eq:int_ic_exact}). In this case, for $N \to \infty$ we get regime (R2') for $\gamma > 3$, while for $\gamma \in (2, 3)$ we get
\begin{align}
\label{eq:l4_plgT}
\nonumber \lim_{N \to \infty} \frac{\eta_N ^{\gamma-1}}{q_N^2}  r_\textnormal{ic}(t) & =g_\tau(t) \Gamma{(3-\gamma)}(\gamma-1)^{3-2(\gamma-1)}\\
 & \times (\gamma-2)^{2(\gamma-1)} \frac{\Gamma{(t+\gamma-1)}}{2\Gamma{(t+2)}},
\end{align}
which is similar to (R3').

\subsection{Intercontact distribution in the configuration model}
\label{sec:intercontact_cm}

The intercontact distribution in the configuration model for large $N$ can be obtained by letting $T \to \infty$ in the corresponding finite-$T$ relations. Let us consider again the case of power-law distributed expected degrees. Since $T \to \infty$, only regimes (R2') and (R3') from Sec.~\ref{sec:intercontact_non_finite} are relevant. For $\gamma > 3$, $r_\textnormal{ic}(t)$ is obtained by letting $T \to \infty$ in~(\ref{eq:l2_uncon_ic}),
\begin{align}
\label{eq:ex3a_cm_ic}
r_\textnormal{ic}(t) \approx g_\tau(t) \left[\frac{\bar{k}}{N} \frac{(\gamma-2)^2}{(\gamma-1)(\gamma-3)}\right]^2.
\end{align}
For $\gamma \in (2, 3)$, $r_\textnormal{ic}(t)$ is obtained by letting $T \to \infty$ in~(\ref{eq:l3_pl_ic}),
\begin{align}
\label{eq:l3_pl_cm_ic}
\nonumber r_\textnormal{ic}(t) & \approx g_\tau(t) \Gamma{(3-\gamma)} (N/\bar{k})^{-(\gamma-1)}\ln{\frac{N (\gamma-1)^2}{\bar{k}(\gamma-2)^2}}\\
 & \times (\gamma-1)^{2-2(\gamma-1)} (\gamma-2)^{2(\gamma-1)}\frac{\Gamma{(t+\gamma-1)}}{\Gamma{(t+2)}}.
\end{align}
Therefore, for $\gamma \in (2,3)$, $P_\textnormal{ic}(t) \propto t^{-(3-\gamma)}$ for $\gamma \ll t \ll \tau$, while for $\gamma > 3$, $P_\textnormal{ic}(t) \propto g_\tau(t)$. 
The exact relation for $r_\textnormal{ic}(t)$ for any network size $N$ can be obtained by letting $T \to \infty$ in~(\ref{eq:pl_exact_ic}),
\begin{align}
\label{eq:pl_exact_ic_cm}
r_\textnormal{ic}(t) &=g_\tau(t) \frac{N^t \bar{k}^t (\gamma-1)^2 }{k_0^{2 t} (t+\gamma-1)^2}\\
\nonumber &\times {}_3 F_{2}\left[t+2, t+\gamma-1, t+\gamma-1; t+\gamma, t+\gamma;  -\frac{N \bar{k}}{k_0^2}\right].
\end{align}
The above relation can be also obtained directly by using in~(\ref{eq:p_ic}) the connection probability in~(\ref{eq:p_cm}) (instead of the connection probability in~(\ref{eq:p_s1_alt})), and computing the integral in~(\ref{eq:int_ic_exact}) with the power-law $\rho(k)$.

\subsection{Intercontact distribution in classical random graphs}
\label{sec:intercontact_rgs}

\begin{figure}
\includegraphics[width=3.0in]{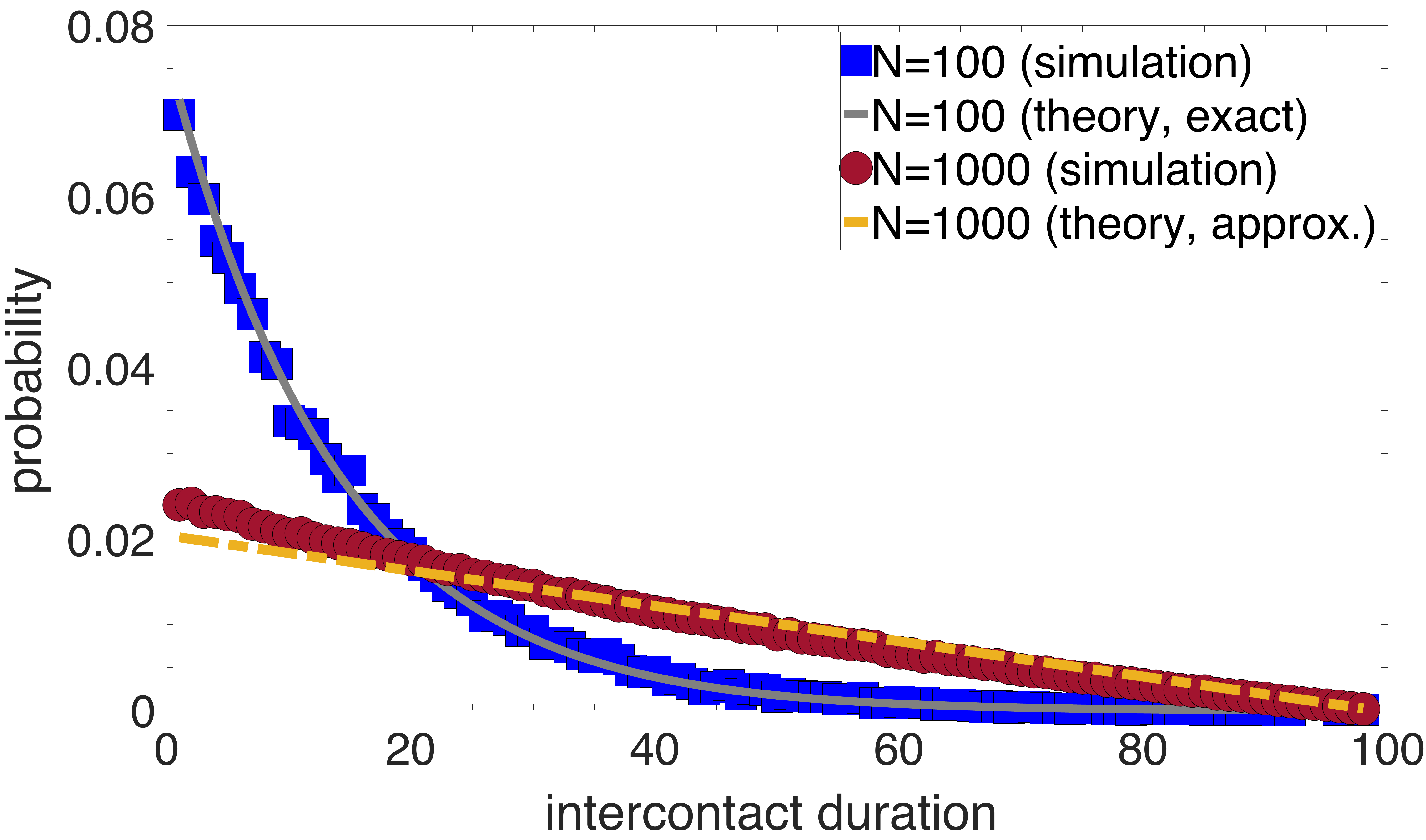}
\caption{Intercontact distribution $P_\textnormal{ic}(t)$ in classical random graphs. The number of nodes $N$ is shown in the legend, the average snapshot degree is $\bar{k}=6$ and $\tau=100$. For $N=100$ the solid line shows the exact theoretical prediction where $r_\textnormal{ic}(t)$ is given by~(\ref{eq:exact_er}). For $N=1000$ the dashed line shows the large-$N$ approximation where $r_\textnormal{ic}(t)$ is given by~(\ref{eq:thermo_ic_rg}).
\label{fig:validation_rg}}
\end{figure}

Finally, the intercontact distribution in classical random graphs for large $N$ can be obtained by assuming $\rho(k)=\delta(k-\bar{k})$ and letting $T \to \infty$ in~(\ref{eq:l2_uncon_ic}). This yields
\begin{equation}
\label{eq:thermo_ic_rg}
r_\textnormal{ic}(t) \approx g_\tau(t)\left(\frac{\bar{k}}{N}\right)^2.
\end{equation}
We can also directly deduce the above expression from the exact relation for $r_\textnormal{ic}(t)$ in classical random graphs. In classical random graphs~(\ref{eq:p_ic}) becomes
\begin{equation}
\label{eq:exact_er}
r_\textnormal{ic}(t) = g_\tau(t) p^{2} (1-p)^t,
\end{equation}
where $p \approx \bar{k}/N$. The above expression is a product of the linearly-decaying term $g_\tau(t)$ and the exponentially-decaying term $(1-p)^t$. For any finite $t$, $(1-p)^t \approx 1$ for sufficiently large $N$, resulting in~(\ref{eq:thermo_ic_rg}). In other words, for any finite observation interval $\tau$, if the network is sufficiently sparse, then $(1-p)^t \approx 1, \forall t \in [1, \tau]$, and the intercontact distribution displays a linear decay that depends only on $\tau$, $P_\textnormal{ic}(t) \propto g_\tau(t)$. Fig.~\ref{fig:validation_rg} illustrates this point with simulations.

\section{Discussion and conclusion}
\label{sec:conclusion}

For sufficiently large hot RHGs the contact distribution decays as a power law with exponent $2+T > 3$, where $T$ is the network temperature. This result holds irrespective of the form of the expected degree distribution, as long as it has a finite $T^{\textnormal{th}}$ moment. Indeed, even if all nodes have the same expected degree, the contact distribution still decays as a power law with the aforementioned exponent. This observation means that the emergence of this power law is due to the nodes' angular (similarity) coordinates, as more similar (closer along the angular direction) nodes tend to stay connected for a longer period of time. This result also holds in cold RHGs, where the power-law exponent is $2+T \in (2, 3)$~\cite{Papadopoulos2019}. Further, in cold RHGs this result holds as long as the expected node degree is finite, which is always the case in reality. Since cold RHGs are more adequate models for real-world networks, this observation suggests that the origin of broad contact distributions in real systems is primarily due to node similarities, instead of popularities (expected degrees). We note that the similarity distribution does not have to be heterogeneous for a power-law contact distribution to emerge. Indeed, in RHGs nodes are uniformly distributed along the angular similarity direction.

We have also seen that if the expected degree distribution is a power law with exponent $\gamma > 2$ and non-finite $T^{\textnormal{th}}$ moment ($\gamma \leq T+1$), then it shapes the contact distribution, which decays as a power law with exponent $\gamma+1 > 3$. This suggests that the power-law contact distribution in this case is due to the heterogeneity of the expected degrees, and not due to node similarities. Indeed, this result holds even if the nodes' similarities are completely ignored, as is the case in the configuration model that emerges for $T \to \infty$. 

For sufficiently large hot RHGs the intercontact distribution decays as a power law with exponent $2-T \in (0,1)$, if the network's temperature is $T \in (1, 2)$ and the expected degree distribution has a finite $T^{\textnormal{th}}$ moment. This result holds irrespective of the form of the expected degree distribution, which means that the emergence of this power law is again due to node similarities. This result also holds in cold RHGs, where the power-law exponent is $2-T \in (1, 2)$, as long as the expected node degree is finite~\cite{Papadopoulos2019}. This observation suggests that the origin of broad intercontact distributions in real networks is also primarily due to node similarities, as more similar nodes tend to re-connect after shorter time intervals.

Finally, we have seen that power-law intercontact distributions with exponent $3-\gamma \in (0,1)$ emerge, if the expected degree distribution is power law with exponent $\gamma \in (2, 3)$ and $\gamma \leq T+1$. In this case, the emergence of the power-law intercontact distribution is due to the heterogeneity of the expected degrees. Indeed, this result holds even if the node similarities are completely ignored, as is the case in the configuration model.

No matter the origin of the power-law intercontact distribution in hot RHGs, our analysis suggests that the exponent of this distribution is always less than one, which is unrealistic. Therefore, hot RHGs are not adequate for modeling real temporal networks, in stark contrast to cold RHGs~\cite{Papadopoulos2019}. At a higher level, our results shed light on the origin of broad (inter)contact distributions in temporal networks, showing that such distributions can emerge both as a result of node similarities and popularities, with the former providing a more probable cause in real-world systems. This observation could aid epidemic control strategies, as it implies that for altering the (inter)contact distributions in real systems one needs to target the (inter)contacts between similar (close along the similarity direction) nodes instead of the (inter)contacts of popular (high expected degree) nodes.

Finally, we note that even though we sometimes presented exact formulas for the (inter)contact distributions (cf.~(\ref{eq:pl_exact}, \ref{eq:pl_exact_ic})), we reach our conclusions by analyzing these distributions in the thermodynamic limit ($N \to \infty$). Our analytical results at this limit can be used as approximations in sufficiently sparse networks, as illustrated by simulations. How much ``sufficiently" sparse a network should be for the approximations to be accurate depends on the network's characteristics, such as its expected degree distribution, how close its parameters are to the boundaries of the different regimes, etc.  Also, in the case of the intercontact distribution in hot RHGs, the value of the observation interval $\tau$ plays a significant role. Indeed, as we have seen in classical random graphs, the approximation in~(\ref{eq:thermo_ic_rg}) can be reached from the exact relation in (\ref{eq:exact_er}) if we assume $(1-p)^t \approx 1$. The larger the $t$ the smaller should the $p \approx \bar{k}/N$ be for the last approximation to hold. In other words, the larger the observation interval $\tau$ the sparser should the network be for the approximation in~(\ref{eq:thermo_ic_rg}) to hold for all $t \in [1, \tau-2]$. Investigating the accuracy of the thermodynamic-limit approximations as a function of network sparsity is beyond the scope of this paper.

\section*{Acknowledgements}
The authors acknowledge support by the TV-HGGs project (OPPORTUNITY/0916/ERC-CoG/0003), co-funded by the European Regional Development Fund and the Republic of Cyprus through the Research and Innovation Foundation.

\appendix

\section{Exchanging the order of the limit with the integral in Eqs.~(\ref{eq:l1_uncon_tmp}) and (\ref{eq:l2_uncon_tmp})}
\label{sec:app1}

We first recall the dominated convergence theorem (DCT), which allows us to exchange the order of limits with integrals, cf.~\cite{goodner1966}. The DCT states that if a function $f_N(x)$ converges to a function $f(x)$ for $N \to \infty$ and $|f_N(x)| \leq h(x), \forall N, x$, where the dominating function $h(x)$ is integrable, $\int h(x)  \mathrm{d} x < \infty$, then $\lim_{N \to \infty} \int f_N(x) \mathrm{d} x=\int \lim_{N \to \infty}  f_N(x) \mathrm{d} x=\int f(x) \mathrm{d} x$.

Now, let 
\begin{equation}
\label{eq:f_N}
f_N(t; k, k') \equiv N^T r_{\textnormal{c}}(t; k, k'),
\end{equation}
where $r_{\textnormal{c}}(t; k, k')$ is given by~(\ref{eq:full_contact_integral}). The integral $I_t \equiv \int \int \lim_{N \to \infty} f_N (t; k, k')  \rho(k) \rho(k') \mathrm{d} k \mathrm{d} k'$ is given by~(\ref{eq:l1_uncon_tmp}) for $t > T$, and it is proportional to $(\overline{k^T})^2$. If $\overline{k^T} \to \infty$, then $I_t \to \infty$. In this case the DCT does not apply and we cannot exchange the order of the limit with the integral in the first line of~(\ref{eq:l1_uncon_tmp}). In other words, the finiteness of $\overline{k^T}$ is a necessary condition for performing this exchange. Similarly, let
\begin{equation}
\label{eq:f_N_tilde}
\tilde{f}_N(t; k, k') \equiv N^t r_{\textnormal{c}}(t; k, k').
\end{equation}
The integral $\tilde{I}_t \equiv \int \int \lim_{N \to \infty} \tilde{f}_N (t; k, k') \rho(k) \rho(k') \mathrm{d} k \mathrm{d} k'$ is given by~(\ref{eq:l2_uncon_tmp}) for $t < T$, and it is proportional to $(\overline{k^t})^2$. If $\overline{k^t} \to \infty$, then $\tilde{I}_t \to \infty$. Thus, for exchanging the order of the limit with the integral in~(\ref{eq:l2_uncon_tmp}) a necessary condition is that $\overline{k^t}$ is finite. Since $t < T$, this condition is satisfied if $\overline{k^T}$ is finite, since all lower-order moments $\overline{k^t}$, $t < T$, will also be finite. Below, we show that the finiteness of $\overline{k^T}$ is also a sufficient condition, i.e., if $\overline{k^T}$ is finite, then the DCT applies for all $ t \geq 1$, and we can exchange the order of the limit with the integral in~(\ref{eq:l1_uncon_tmp}) and (\ref{eq:l2_uncon_tmp}).

We first recall the following facts about the hypergeometric function (see~\cite{special_functions_book} for further details). 
The hypergeometric function is defined by the Gauss series
\begin{equation}
\label{eq:hypeq}
{}_2 F_{1}[a, b; c; z]=\sum_{n=0}^\infty \frac{(a)_n (b)_n}{(c)_n} \frac{z^n}{n!},
\end{equation}
for $|z| < 1$, and by analytic continuation elsewhere. The symbol $(q)_n$ is the  Pochhammer symbol, defined as $(q)_n=1$ for $n=0$, and $(q)_n=q (q+1) \ldots (q+n-1)$ for $n > 0$.
For $z \geq 0$ the following identities hold (Pfaff's transformations):
\begin{align}
\label{eq:pfaff1}
{}_2 F_{1}[a, b; c; -z] &= (1+z)^{-a} {}_2 F_{1}[a, c-b; c; \frac{z}{z+1}]\\
\label{eq:pfaff2}
&= (1+z)^{-b} {}_2 F_{1}[c-a, b; c; \frac{z}{z+1}].
\end{align}
Further, it is easy to see from~(\ref{eq:hypeq}) that for positive $a, b, c$ and $z \in [0,1]$, the following upper bound holds:
\begin{equation}
\label{eq:obs1}
{}_2 F_{1}[a, b; c; z] \leq {}_2 F_{1}[a, b; c; 1],
\end{equation}
where ${}_2 F_{1}[a, b; c; 1]$ is given by~(\ref{eq:hypeq}) for $z=1$. If $c-a-b > 0$, then 
\begin{equation}
\label{eq:limobs1}
{}_2 F_{1}[a, b; c; 1]=\frac{\Gamma{(c)}\Gamma{(c-a-b)}}{\Gamma{(c-a)}\Gamma{(c-b)}}.
\end{equation}
Finally, we can see that for $z \in [0,1]$ and $m \geq 0$, the following inequality holds: ${}_2 F_{1}[-m, b; b+1; 1] \leq {}_2 F_{1}[-m, b; b+1; z] \leq {}_2 F_{1}[-m, b; b+1; 0]$, meaning that
\begin{equation}
\label{eq:obs2}
\frac{\Gamma{(b+1)}\Gamma{(m+1)}}{\Gamma{(b+m+1)}}\leq {}_2 F_{1}[-m, b; b+1; z] \leq 1.
\end{equation} 

Now, let  
\begin{equation}
\label{eq:z}
z_{ij} \equiv \frac{N \bar{k}}{(1-1/T) k_i k_j}. 
\end{equation}
Using~(\ref{eq:pfaff1}) and~(\ref{eq:pfaff2}), we can re-write~(\ref{eq:full_contact_integral}) as
\begin{align}
\label{eq:rc_alt_1}
r_\textnormal{c}(t ; k_i, k_j ) &= g_\tau(t)\frac{T}{T+2} z_{ij}^2 (1+z_{ij})^{-(t+2)}\\
\nonumber & \times {}_2 F_{1}[t+2, 1; T+3; \frac{z_{ij}}{z_{ij}+1}]\\
\label{eq:rc_alt_2}
&= g_\tau(t)\frac{T}{T+2} z_{ij}^2 (1+z_{ij})^{-(T+2)}\\
\nonumber &\times {}_2 F_{1}[T-t+1, T+2; T+3; \frac{z_{ij}}{z_{ij}+1}],
\end{align}
where $z_{ij}$ given by (\ref{eq:z}). We note that $z_{ij} > 0$. Thus, the ratio $z_{ij}/(z_{ij}+1)$ takes values in $(0, 1)$. Using (\ref{eq:rc_alt_1}), (\ref{eq:obs1}) and (\ref{eq:limobs1}), we can write the following upper bound for $\tilde{f}_N(t; k, k')$ in (\ref{eq:f_N_tilde}), for $t < T$:
\begin{align}
\label{eq:h1}
\nonumber \tilde{f}_N(t; k_i, k_j) & = N^t g_\tau(t) \frac{T}{T+2} z_{ij}^{2} (1+z_{ij})^{-(t+2)}\\
\nonumber & \times {}_2 F_{1}[t+2, 1; T+3; \frac{z_{ij}}{z_{ij}+1}]\\
\nonumber & \leq g_\tau(t) \frac{T}{T+2} \frac{(1-1/T)^t}{\bar{k}^t} k_i^t k_j^t\\
\nonumber & \times {}_2 F_{1}[t+2, 1; T+3; 1]\\
\nonumber &= g_\tau(t) \frac{T}{T-t}  \frac{(1-1/T)^t}{\bar{k}^t} k_i^t k_j^t\\
&\equiv h_1 (t; k_i, k_j).
\end{align}
We note that the upper bound $h_1 (t; k_i, k_j)$ is the $\lim_{N\to\infty} \tilde{f}_N(t; k_i, k_j)$, obtained in~(\ref{eq:l2}). If $\overline{k^T}$ is finite, then the dominating function $h_1(t; k, k') \rho(k) \rho(k')$ is integrable for $t < T$, $\int \int h_1 (t; k, k') \rho(k) \rho(k')  \mathrm{d} k  \mathrm{d} k' \propto (\overline{k^{t}})^2$. Therefore, the DCT applies and we can exchange the order of the limit with the integral in~(\ref{eq:l2_uncon_tmp}).

Moreover, using (\ref{eq:rc_alt_2}) and the upper bound in (\ref{eq:obs2}), we can write the following upper bound for $f_N(t; k, k')$ in (\ref{eq:f_N}), for $t \geq T + 1$:
\begin{align}
\label{eq:h2}
\nonumber f_N(t; k_i, k_j) & = N^T g_\tau(t)\frac{T}{T+2} z_{ij}^2 (1+z_{ij})^{-(T+2)}\\
\nonumber &\times {}_2 F_{1}[T-t+1, T+2; T+3; \frac{z_{ij}}{z_{ij}+1}] \\
\nonumber & <  g_\tau(t)\frac{T}{T+2} \frac{(1-1/T)^T}{\bar{k}^T} k_i^T k_j^T\\\
& \equiv h_2 (t; k_i, k_j).
\end{align}
The dominating function $h_2(t; k, k') \rho(k) \rho(k')$ is again integrable if $\overline{k^T}$ is finite, $\int \int h_2 (t; k, k') \rho(k) \rho(k') \mathrm{d} k  \mathrm{d} k' \propto (\overline{k^{T}})^2$. Therefore, we can exchange the order of the limit with the integral in~(\ref{eq:l1_uncon_tmp}) for all $t \geq T+1$.

For the case $T < t < T+1$, we can use (\ref{eq:rc_alt_2}), (\ref{eq:obs1}) and (\ref{eq:limobs1}), to write the following upper bound for $f_N(t; k, k')$:
\begin{align}
\label{eq:h3}
\nonumber f_N(t; k_i, k_j) & = N^T g_\tau(t)\frac{T}{T+2} z_{ij}^2 (1+z_{ij})^{-(T+2)}\\
\nonumber &\times {}_2 F_{1}[T-t+1, T+2; T+3; \frac{z_{ij}}{z_{ij}+1}]\\
\nonumber & \leq g_\tau(t)\frac{T}{T+2} \frac{(1-1/T)^T}{\bar{k}^T} k_i^T k_j^T\\
\nonumber & \times {}_2 F_{1}[T-t+1, T+2; T+3; 1]\\
\nonumber & = g_\tau(t)\frac{T  \Gamma{(T+2)} (1-1/T)^T}{\bar{k}^T}\frac{\Gamma{(t-T)}}{\Gamma{(t+2)}} k_i^T k_j^T\\
& \equiv h_3 (t; k_i, k_j).
\end{align}
$h_3 (t; k_i, k_j)$ is the $\lim_{N\to\infty} f_N(t; k_i, k_j)$, obtained in~(\ref{eq:l1}). The dominating function $h_3(t; k, k') \rho(k) \rho(k')$ is integrable if $\overline{k^T}$ is finite. Therefore, we can exchange the order of the limit with the integral in~(\ref{eq:l1_uncon_tmp}) for all $t > T$.

\section{Exchanging the order of the limit with the integral in Eqs.~(\ref{eq:l1_uncon_ic_tmp}) and (\ref{eq:l2_uncon_ic_tmp})}
\label{sec:app2}

We follow similar steps as in Appendix~\ref{sec:app1}. Let
\begin{equation}
\label{eq:phi_N}
\phi_N(t; k, k') \equiv N^T r_{\textnormal{ic}}(t; k, k'),
\end{equation}
and
\begin{equation}
\label{eq:phi_tilde}
\tilde{\phi}_N(t; k, k') \equiv N^2 r_{\textnormal{ic}}(t; k, k'),
\end{equation}
where  $r_{\textnormal{ic}}(t; k, k')$ is given by~(\ref{eq:full_intercontact_integral}). 

The integral $I'_t \equiv \int \int \lim_{N \to \infty} \phi_N (t; k, k') \rho(k) \rho(k') \mathrm{d} k \mathrm{d} k'$ is given by~(\ref{eq:l1_uncon_ic_tmp}) for $T \in (1, 2)$, and it is proportional to $(\overline{k^T})^2$. If $\overline{k^T} \to \infty$, $I'_t \to \infty$. In this case the DCT does not apply and we cannot exchange the order of the limit with the integral in the first line of~(\ref{eq:l1_uncon_ic_tmp}). Thus, the finiteness of $\overline{k^T}$ is a necessary condition for performing this exchange. Similarly, the integral $\tilde{I'}_t \equiv \int \int \lim_{N \to \infty} \tilde{\phi}_N (t; k, k') \rho(k) \rho(k') \mathrm{d} k \mathrm{d} k'$ is given by~(\ref{eq:l2_uncon_ic_tmp}) for $T > 2$, and it is proportional to $(\overline{k^2})^2$. If $\overline{k^2} \to \infty$, $\tilde{I'}_t \to \infty$. Therefore, a necessary condition for exchanging the order of the limit with the integral in~(\ref{eq:l2_uncon_ic_tmp}) is that $\overline{k^2}$ is finite. Below, we show that the above conditions are also sufficient in each case.

Using~(\ref{eq:pfaff1}) and (\ref{eq:pfaff2}), we can re-write~(\ref{eq:full_intercontact_integral}) as
\begin{align}
\label{eq:ric_alt_1}
r_\textnormal{ic}(t ; k_i, k_j ) &= g_\tau(t)\frac{T}{t+T} z_{ij}^t (1+z_{ij})^{-(t+2)}\\
\nonumber & \times {}_2 F_{1}[t+2, 1; t+T+1; \frac{z_{ij}}{z_{ij}+1}]\\
\label{eq:ric_alt_2}
&= g_\tau(t)\frac{T}{t+T} z_{ij}^t (1+z_{ij})^{-(t+T)}\\
\nonumber &\times {}_2 F_{1}[T-1, t+T; t+T+1; \frac{z_{ij}}{z_{ij}+1}],
\end{align}
where $z_{ij}$ is given by~(\ref{eq:z}).

Using~(\ref{eq:ric_alt_2}), (\ref{eq:obs1}) and (\ref{eq:limobs1}), we can write the following upper bound for $\phi_N(t; k, k')$ in~(\ref{eq:phi_N}), for $T \in (1, 2)$:
\begin{align}
\label{eq:h4}
\nonumber \phi_N(t; k_i, k_j) & = N^T g_\tau(t)\frac{T}{t+T} z_{ij}^t (1+z_{ij})^{-(t+T)}\\
\nonumber &\times {}_2 F_{1}[T-1, t+T; t+T+1; \frac{z_{ij}}{z_{ij}+1}]\\\
\nonumber & \leq g_\tau(t)\frac{T}{t+T}  \frac{(1-1/T)^T}{\bar{k}^T} k_i^T k_j^T\\
\nonumber &\times {}_2 F_{1}[T-1, t+T; t+T+1; 1]\\
\nonumber &= g_\tau(t) \frac{T \Gamma{(2-T)} (1-1/T)^T}{\bar{k}^T} \frac{\Gamma{(t+T)}}{\Gamma{(t+2)}}k_i^T k_j^T\\
&\equiv h_4 (t; k_i, k_j).
\end{align}
We note that the upper bound $h_4(t; k_i, k_j)$ is the $\lim_{N\to\infty} \phi_N(t; k_i, k_j)$, obtained in~(\ref{eq:l1_ic}). The dominating function $h_4(t; k, k') \rho(k) \rho(k')$ is integrable if $\overline{k^T}$ is finite, $\int \int h_4 (t; k, k')  \rho(k) \rho(k') \mathrm{d} k  \mathrm{d} k' \propto  (\overline{k^{T}})^2$. Therefore, the DCT applies and we can exchange the order of the limit with the integral in~(\ref{eq:l1_uncon_ic_tmp}).

Moreover, using (\ref{eq:ric_alt_1}), (\ref{eq:obs1}) and (\ref{eq:limobs1}), we can write the following upper bound for $\tilde{\phi}_N(t; k, k')$ in~(\ref{eq:phi_tilde}), for $T > 2$:
\begin{align}
\label{eq:h4}
\nonumber \tilde{\phi}_N(t; k_i, k_j) & = N^2 g_\tau(t)\frac{T}{t+T} z_{ij}^t (1+z_{ij})^{-(t+2)}\\
\nonumber & \times {}_2 F_{1}[t+2, 1; t+T+1; \frac{z_{ij}}{z_{ij}+1}]\\
\nonumber & \leq g_\tau(t)\frac{T}{t+T}  \frac{(1-1/T)^2}{\bar{k}^2} k_i^2 k_j^2\\
\nonumber & \times {}_2 F_{1}[t+2, 1; t+T+1; 1]\\
\nonumber &=  g_\tau(t)\frac{T}{T-2} \frac{(1-1/T)^2}{\bar{k}^2} k_i^2 k_j^2\\
&\equiv h_5 (t; k_i, k_j).
\end{align}
$h_5(t; k_i, k_j)$ is the $\lim_{N\to\infty} \tilde{\phi}_N(t; k_i, k_j)$, obtained in~(\ref{eq:l2_ic}). The function $h_5(t; k, k') \rho(k) \rho(k')$ is integrable if $\overline{k^2}$ is finite, $\int \int h_5 (t; k, k') \rho(k) \rho(k') \mathrm{d} k  \mathrm{d} k' \propto  (\overline{k^{2}})^2$, and we can exchange the order of the limit with the integral in~(\ref{eq:l2_uncon_ic_tmp}).

%

\end{document}